\documentclass[twocolumn]{aastex63}

\newcommand{\chandra}{\textit{Chandra}}
\newcommand{\rxte}{\textit{RXTE}}

\newcommand{\sax}{\textit{BeppoSAX}}

\newcommand{\nustar}{\textit{NuSTAR}}

\newcommand\cl{CL 0217+70}

\usepackage{multirow}
\usepackage{wrapfig} 
\usepackage{lineno}
\received{June 18, 2022}
\revised{October 26, 2022}
\accepted{November 8, 2022}
\shorttitle{The \nustar~and \chandra~view of \cl~}
\shortauthors{T\"{u}mer et al.}
\begin{document}

\title{The \nustar~and \chandra~view of \cl~and Its Tell-Tale Radio Halo}

\correspondingauthor{Ay\c{s}eg\"{u}l T\"{u}mer}
\email{aysegultumer@gmail.com}

\author[0000-0002-3132-8776]{Ay\c{s}eg\"{u}l T\"{u}mer}
\affiliation{Department of Physics \& Astronomy, The University of Utah, 115 South 1400 East, Salt Lake City, UT 84112, USA}
\affiliation{Kavli Institute for Astrophysics and Space Research, Massachusetts Institute of Technology, 77 Massachusetts Avenue, Cambridge, MA 02139, USA}

\author[0000-0001-9110-2245]{Daniel R. Wik}
\affiliation{Department of Physics \& Astronomy, The University of Utah, 115 South 1400 East, Salt Lake City, UT 84112, USA}

\author[0000-0001-9110-2245]{Xiaoyuan Zhang}
\affiliation{Leiden Observatory, Leiden University, PO Box 9513, 2300 RA Leiden, The Netherlands}
\affiliation{SRON Netherlands Institute for Space Research, Sorbonnelaan 2, 3584 CA Utrecht, The Netherlands}

\author[0000-0002-8286-646X]{Duy N. Hoang}
\affiliation{Hamburger Sternwarte, University of Hamburg, Gojenbergsweg 112, 21029 Hamburg, Germany}

\author[0000-0003-2754-9258]{Massimo Gaspari}
\affiliation{Department of Astrophysical Sciences, Princeton University, 4 Ivy Lane, Princeton, NJ 08544-1001, USA}

\author[0000-0002-0587-1660]{Reinout J. van Weeren}
\affiliation{Leiden Observatory, Leiden University, PO Box 9513, 2300 RA Leiden, The Netherlands}

\author[0000-0001-5636-7213]{Lawrence Rudnick}
\affiliation{Minnesota Institute for Astrophysics, University of Minnesota, 116 Church St. S.E., Minneapolis, MN 55455, USA}

\author[0000-0003-1619-3479]{Chiara Stuardi}
\affiliation{Dipartimento di Fisica e Astronomia, Universit\`a di Bologna, via Gobetti 93/2, I-40129 Bologna, Italy}
\affiliation{INAF - Istituto di Radioastronomia di Bologna, Via Gobetti 101, I-40129 Bologna, Italy}

\author[0000-0002-7031-4772]{Fran\c{c}ois Mernier}
\affiliation{European Space Agency (ESA), European Space Research and Technology Centre (ESTEC), Ke- plerlaan 1, 2201 AZ Noordwijk, The Netherlands}

\author[0000-0002-9714-3862]{Aurora Simionescu}
\affiliation{SRON Netherlands Institute for Space Research, Sorbonnelaan 2, 3584 CA Utrecht, The Netherlands}
\affiliation{Leiden Observatory, Leiden University, PO Box 9513, NL-2300 RA Leiden, The Netherlands}
\affiliation{Netherlands Institute for Radio Astronomy (ASTRON), P.O. Box 2, 7990 AA Dwingeloo, The Netherlands}

\author [0000-0002-8882-6426]{Randall A. Rojas Bolivar}
\affiliation{Department of Physics \& Astronomy, The University of Utah, 115 South 1400 East, Salt Lake City, UT 84112, USA}

\author[0000-0002-0765-0511]{Ralph Kraft}
\affiliation{Harvard-Smithsonian Center for Astrophysics, 60 Garden Street, Cambridge, MA 02138, USA}

\author[0000-0003-1949-7005]{Hiroki Akamatsu}
\affiliation{SRON Netherlands Institute for Space Research, Sorbonnelaan 2, 3584 CA Utrecht, The Netherlands}

\author[0000-0002-2697-7106]{Jelle de Plaa}
\affiliation{SRON Netherlands Institute for Space Research, Sorbonnelaan 2, 3584 CA Utrecht, The Netherlands}

%% Mark off the abstract in the ``abstract'' environment. 
\begin{abstract}

Mergers of galaxy clusters are the most energetic events in the Universe, driving shock and cold fronts, generating turbulence, and accelerating particles that create radio halos and relics. The galaxy cluster \cl~is a remarkable late stage merger, with a double peripheral radio relic and a giant radio halo. A \chandra~study detects surface brightness edges that correspond to radio features within the halo. In this work, we present a study of this cluster with \nustar\ and \chandra\ data using spectro-imaging methods. The global temperature is found to be \textit{kT} = 9.1 keV. We set an upper limit for the IC flux of $\sim$2.7$\times$10$^{-12}$~erg~s$^{-1}$~cm$^{-2}$, and a lower limit to the magnetic field of 0.08~$\mu$G. Our local IC search revealed a possibility that IC emission may have a significant contribution at the outskirts of the radio halo emission and on/near shock regions within $\sim$0.6 r$_{500}$ of clusters. We detected a ``hot spot" feature in our temperature map coincident with a surface brightness edge, but our investigation on its origin is inconclusive. If the ``hot spot" is the downstream of a shock, we set a lower limit of \textit{kT} $>$ 21 keV to the plasma, that corresponds to $\mathcal{M}$ $\sim$ 2. We found three shock fronts within 0.5~r$_{500}$. Multiple weak shocks within the cluster center hint at an ongoing merger activity and continued feeding of the giant radio halo. \cl~ is the only example hosting these secondary shocks in multiple form.
\end{abstract}


\section{Introduction} \label{sec:intro}

Galaxy clusters are the largest gravitationally bound structures in the universe, which contain hundreds of galaxies within a radius of 1–2 Mpc \citep{abell58}. The intracluster medium (ICM) is an optically thin hot plasma ($\sim10^{7}-10^{8}$ K) in X-rays, that fills in between the galaxies inside galaxy clusters.

Mergers between galaxy clusters are the most energetic (10$^{63}$-10$^{64}$~ergs) events in the universe since the Big Bang. These mergers drive shocks that heat and mix the thermal gas while also (re)accelerating electrons and cosmic rays through Fermi-like acceleration processes across shocks and turbulent eddies \citep{brunetti14}. Turbulence is a common phenomenon in the ICM, recurrently driven by mergers at large scales (e.g., \citealt{gaspari13b,eckert17}), sustained by AGN feedback (e.g., \citealt{gaspari15,hofmann16,yang19,wittor20}) and ram-pressure stripping processes at small scales (e.g., \citealt{degrandi16,clavico19}).

The ICM carries the X-ray traces of mergers such as shock fronts that are detected as spatial variations in the temperature and density \citep[see, e.g.][and references therein]{markevitch99}. While surface brightness (SB) edges hint at the existence of a shock or cold front, temperature measurements at both sides of the SB edges can distinguish these two scenarios. Shocks caused by mergers have relatively low Mach numbers, $\mathcal{M}$ $\lesssim$ 3 \citep[see, e.g.][]{gabici03,ryu03}.

Mergers also produce relativistic particles emitting via non-thermal synchrotron process that can be detected by radio observations. The diffuse radio emission can be in the form of radio relics and radio halos. Relics are defined as extended radio emission features that are not associated with an active cluster radio galaxy or central region of a galaxy cluster \citep{giovannini04}. They are located at the peripheries of clusters and are characterized by polarized radio emission with steep spectra and elongated shapes \citep{ensslin98}. So far, $\sim$45 clusters are known to host these structures and almost as many candidates \citep{weeren19}. Double radio relic systems, where relics appear on opposing sides of the cluster, are even rarer. 17 such systems are known to date and only 7 of them host giant radio halos \citep{weeren19}. 

Radio halos are Mpc size structures that are found at the center of dynamically disturbed clusters with large scale particle acceleration processes \citep{cassano10,feretti12}, yet they are not ubiquitous \citep{schellenberger19}. While evidence suggests that radio relics trace shock fronts from the initial collisions of clusters, the origin of radio halos is less clear \citep[see, e.g.]{brunetti12}. The prevailing view is that the relativistic electrons producing radio halos are accelerated by turbulence in the ICM. 

Unlike relics, most radio halos have no known shock front-radio emission edge connection except for a few cases (e.g., Abell 520 \citep{hoang19}, the Toothbrush \citep{weeren16}, the Coma Cluster \citep{brown11,planck13a}, the Bullet Cluster \citep{shimwell14}, and Abell 2146 \citep{hlavacek17}). This small number of cluster mergers with known shock front-radio edge connection suggests that turbulence formed directly behind shocks is what accelerates the electrons in radio halos, whether from the initial collision or a subsequent settling of the ICM. The Bullet cluster may be an example of the former case (initial collision), since its radio halo extends right up to the bow shock—driven by the subcluster’s first core passage \citep{shimwell14}. In contrast, the Coma cluster seems to be the only example of the latter scenario (subsequent settling of the ICM) so far. Recent eROSITA images, combined with SZ and radio data, convincingly illustrate the formation and connection to a radio halo edge of a secondary shock \citep{zhangcong20}. This secondary shock is also detected in X-rays by \citet{simionescu13}.

\begin{deluxetable*}{lccccc}
\tabletypesize{\scriptsize}
\tablecaption{Observation log. \label{tab:obslog}}
\tablehead{
\colhead{} &\colhead{Observation ID} & \colhead{Start Date} & \colhead{Equatorial coordinates} & {Effective}\\[-0.5em]
\colhead{} & \colhead{} & \colhead{(YYYY-mm-dd)}& \colhead{(J2000)} & \colhead{exposure (ks)}
}
\startdata
\nustar~ & 70701001002 & 2021-07-04 & 02:16:58.2 	+70:36:48 & $\sim$230 \\
& 70701001004 & 2021-07-08 &02:17:06.9 	+70:37:05 & $\sim$106 \\
\chandra~& 16293 & 2014-12-01 & 02:16:49.00 +70:35:52.00 & $\sim$25\\
\enddata
\end{deluxetable*}

In addition, the same relativistic electrons producing the synchrotron-powered radio halo also upscatter  cosmic microwave background (CMB) photons to X-ray energies. The ratio of fluxes between the synchrotron and this inverse Compton (IC) emission is simply the ratio of the energy densities of their respective radiation fields: the magnetic field strength $B$ and the CMB, respectively \citep{wik14}. Since the latter is very well known, an IC detection or upper limit directly leads to an estimate or lower limit on the volume-averaged value of $B$, a quantity that is poorly constrained in galaxy clusters generally.

Past \nustar\ searches for IC emission \citep[see, e.g.][]{wik11,wik14,cova19,rojas21} constrain $B\ga0.1-0.2~\mu$G comparable to the strength estimated from IC detections made with \rxte~ and \sax. Those detections remain controversial, but if IC emission could be definitively measured in a cluster, implying $B\sim0.1~\mu$G, the dynamical role of magnetic fields in clusters would be confirmed to be negligible, as is currently assumed in simulations and mass scaling relations used in cluster cosmology \citep[see, e.g.][]{vikhlinin09}).

One of these rare double radio relic system at \textit{z}~=~0.18 \citep{zhang20} is \cl. This cluster hosts the most extended (3.5 Mpc) radio relic known accompanied by a giant radio halo with the largest projected distance of 1.5~-~1.8 Mpc \citep{brown11b,hoang21}. Its two radio relics are mostly separated from its radio halo and lie at a large projected distance, indicating the first core passage was not recent. Intriguingly, two X-ray SB discontinuities or edges have recently been reported from a 25~ks \chandra\ observation \citep{zhang20}, which are roughly coincident with spectral edges in the radio halo, i.e., locations where the spectral index suddenly changes from flat to steep, suggesting a transition to a zone of particle acceleration \citep{hoang21}. However, the nature of these X-ray SB edges are unknown. A confirmation that at least one of these edges is a shock front would add \cl~to a rare class of objects with the potential to unravel how particles in radio halos are accelerated.

In addition, the \chandra\ image of \cl\ exhibits a 1\arcmin-wide SB depression, or ``Channel" \citep{zhang20}. Such structures can only exist if they are supported by some form of non-thermal pressure: a locally enhanced value of either $P_B$ or of the density of non-thermal particles. The former case would suppress all X-ray emission, while the latter case would lead to enhanced IC emission, more easily detected at hard X-ray energies.

In this work, we study \cl~using recent \nustar~and archival \chandra~data to study the nature of these SB edges, as well as searching for global and local IC emission.

The paper is organized as follows: observations, data reduction process and the background assessment of the \nustar\ data are presented in Section~\ref{sec:reduction}. In Section~\ref{sec:analysis}, methods used for the various analyses of the cluster and their respective results are presented. We discuss our findings in Section~\ref{sec:discussion}, and present our conclusions in Section~\ref{sec:conclusion}.

Throughout this paper, we assume the $\Lambda$CDM cosmology with {\it H$_{0}$} = 70 km s$^{-1}$ Mpc$^{-1}$, $\Omega_{M}$ = 0.3, $\Omega_{\Lambda}$ = 0.7. According to these assumptions, at the cluster redshift (\textit{z} = 0.18), a projected intracluster distance of 100 kpc corresponds to an angular separation of $\sim$32$\arcsec$. All uncertainties are quoted at the 68\% confidence levels unless otherwise stated.

\section{Observations and data reduction} \label{sec:reduction}
\subsection{\nustar}
In this work, we use \nustar~\citep[Nuclear Spectroscopic Telescope Array;][]{harrison13} observations of \cl\ with Observation IDs 70701001002 and 70701001004 using both focal plane modules, namely, FPMA and FPMB. The specifications of the data used in our analysis are summarized in Table~\ref{tab:obslog}.

\begin{figure*}
\centering
\includegraphics[width=170mm]{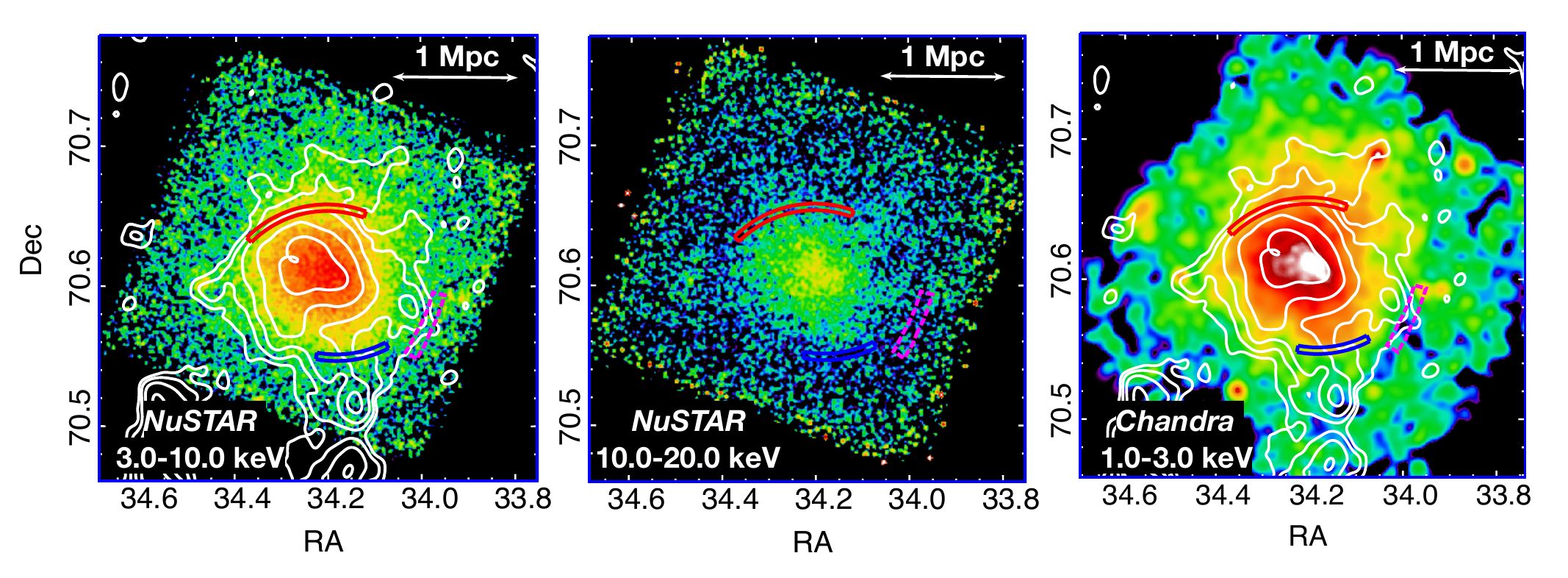}
\caption{Background subtracted, exposure corrected photon images of \cl~ in the 3.0~-~10.0 keV \nustar~ ({\it left panel}), 10.0~-~20.0 keV \nustar~ ({\it middle panel}), and 1.0~-~3.0 keV \chandra\ ({\it right panel}) bands. LOFAR contour levels (white) correspond to $\sigma_{rms}\times$ [3, 6, 12, 24, 48]  ($\sigma_{rms}$ = 330~$\mu$Jy/beam [46$\arcsec\times$45$\arcsec$]) \citep{hoang21}. Red and blues arcs correspond to the location of northern and southern SB edges proposed by \citet{zhang20}, respectively. The channel region is indicated with dashed magenta rectangle.
\label{fig:nuphoton}}
\end{figure*}

\begin{figure*}
\centering
\includegraphics[width=130mm]{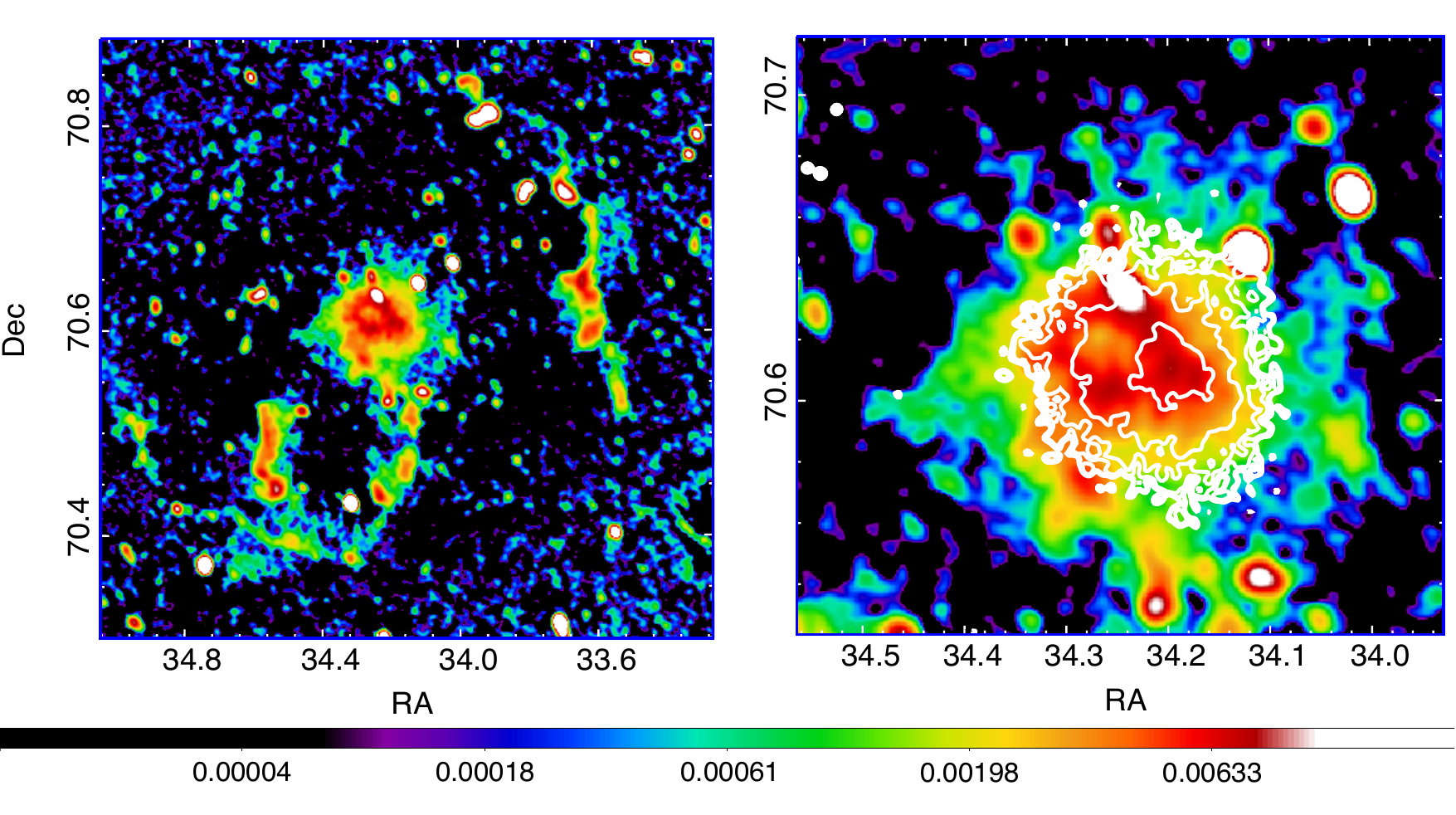}
\caption{LOFAR 26.6$\arcsec$ $\times$ 22.5$\arcsec$ resolution image \citep{hoang21} with 3.0-10.0 keV \nustar~ contours overlaid. \label{fig:lofar}}
\end{figure*}

In order to filter the data, standard pipeline processing using HEASoft (v.~6.28) and NuSTARDAS (v.~2.0.0.) tools are used. To clean the event files, the stage 1 and 2 of the NuSTARDAS pipeline processing script {\tt nupipeline} are used. Regarding the cleaning of the event files for the passages through the South Atlantic Anomaly (SAA) and a ``tentacle"-like region of higher activity near part of the SAA, instead of using SAAMODE=STRICT and TENTACLE=yes calls, we have created light curves and applied different filters to create good time intervals (GTIs) manually without fully discarding the passage intervals.

The new set of GTIs are reprocessed with {\tt nupipeline} stages 1 and 2, and images are generated at different energy bands with XSELECT. To create exposure maps, nuexpomap are used. To produce the corresponding spectra for the regions of interest as well as the corresponding Response Matrix Files (RMFs) and Ancillary Response Files (ARFs), stage 3 of {\tt nuproducts} pipeline are used.

Main components of the \nustar~background are instrument Compton scattered continuum emission, instrument activation and emission lines, cosmic X-ray background from the sky leaking past the aperture stops, reflected solar X-rays, and focused and ghost-ray cosmic X-ray background. Modeling the background where there is a lack of cluster emission regions is not straightforward since the ICM emission becomes an additional component in the background fitting procedure. To apply these models for the cluster background assessment, a set of IDL routines called {\tt nuskybgd} \citep{wik14}, which defines the background spatially and spectrally, is utilized. 

The procedure starts with selecting regions in the FOV, where the cluster emission is inherently present yet not the most dominant. To account for the ICM emission, a single temperature {\tt apec} model is included in the full set of models, and jointly fitted with the background components. The point sources detected in the \chandra~FOV were not originally excluded from these regions, since they did not become apparent until after the background subtraction. While they may bias the focused cosmic X-ray background (fCXB) component in some regions, the average fCXB level is used to produce background spectra in all regions, lessening their impact, which is already only $\sim$10\% of the total background. We discuss the systematics in Section~\ref{sec:bgdfit}.

The global background model is then used to create background images, which are then subtracted from the images and are corrected by the corresponding exposure maps. Background subtracted, exposure corrected images at different energy bands are presented in the left and middle panels of Fig.~\ref{fig:nuphoton}. Region selection and the background fits are presented in Section~\ref{sec:bgdfit}.

Once the background is defined for any region in the FOV both spatially and spectrally, the next step is to select regions of interest, to extract spectra and the corresponding files and generating the specific background model, followed by spectral fitting to evaluate the physical properties of the ICM.

\subsection{\chandra}
We use the observation and blank-sky background event files obtained from the 25~ks \chandra\ archival data (Observation ID: 16293) produced by \citet{zhang20} for the data analysis. We use \chandra~Interactive Analysis of Observations (CIAO) v4.12 package with CALDB 4.9.0 for extracting the spectra. Background subtracted, exposure corrected image in the 1.0~-~3.0 keV band are presented in the right panel of Fig.~\ref{fig:nuphoton}.

In addition to \nustar~and \chandra~data, we also represent the LOFAR data with courtesy of \citet[][Fig. 4]{hoang21}. The LOFAR image of the cluster showing the relics and the central region is presented in Fig.~\ref{fig:lofar}.

\section{Data analysis and results} \label{sec:analysis}

\subsection{Global properties}

We began our analysis with the assessment of the global properties of the cluster. We selected a circular region centered at the X-ray emission peak with \textit{r} = 5.2\arcmin~ ($\sim$1 Mpc), following the extent of the radio halo emission. We extracted a spectrum from \nustar\, which then was fit by a single temperature {\tt apec} model \citep{smith01}, i.e. {\tt constant} $\times$ {\tt apec}, i.e. 1T model, with {\tt XSPEC} (v. 12.11.1). 

The redshift value of the cluster is freed to vary due to the gain issue of \nustar~ (Rojas Bolivar et al., submitted). Once the best-fit values are found for the redshift, we then freeze these parameters during error calculations. We apply the maximum likelihood-based statistic (hereafter, C-stat) appropriate for Poisson data as proposed by \citet{cash79}. Photon counts used in spectral analysis are grouped to have at least 3 counts in each bin. 

We then applied this method to the joint analysis of \nustar~and \chandra~data. For all of the joint analyses, we convolve the model with {\tt TBabs} since \chandra~is susceptible to \textit{N$_{H}$}, and we let this parameter free. \cl\ lies within 9\arcdeg\ of the Galactic plane and consequently suffers from a high foreground column density ($N_H \sim 8\times10^{21}$~cm$^{-2}$), which absorbs most X-ray photons with energies below a few keV. Near the plane, it is also expected that the column density will be variable on arcminute scales, which cannot be reliably predicted from HI estimates. Due to \nustar's lack of bandpass sensitivity below 3 keV, we are able to the ignore foreground absorption at this level \citep{rojas21} for \nustar~analysis for simplicity. However, we report $\sim$0.2 keV difference with/without the inclusion of an absorption model with $N_H \sim 8\times10^{21}$~cm$^{-2}$ for the global fit. However, when the N$_{H}$ parameter is freed to vary, the face value becomes 10$^{17}$~cm$^{-2}$ where the lower limit hits the hard limit of zero with the higher limit reaching on to 10$^{19}$~cm$^{-2}$. This unphysical behavior of the parameter points to NuSTAR's lack of sensitivity to N$_{H}$ at low energies.

\begin{deluxetable}{lccc}
\tabletypesize{\scriptsize}
\tablewidth{0pt} 
\tablecaption{Spectral parameters and 1$\sigma$ uncertainty ranges of \nustar\ global spectrum fits in 3.0~-~20.0 band for 1T: {\tt constant} $\times$ {\tt apec}, 2T: {\tt constant} $\times$ ({\tt apec} + {\tt apec}) and 1T + IC: {\tt constant} $\times$ ({\tt apec} + {\tt powerlaw}). For the 2T model, abundance and redshift values of two {\tt apec} components are tied to each other within instruments. {\tt apec} normalization (\textit{n}) is given in $\frac{10^{-14}}{4\pi \left [ D_A(1+z) \right ]^2}\int n_en_HdV$  where {\tt powerlaw} normalization ($\kappa$) is \textit{photons}  $\textit{keV}^{-1}~cm^{-2}~s^{-1}$ at 1 keV.
\label{tab:globalfit}}
\tablehead{\\[-0.95em]
\colhead{}& \colhead{1T} &\colhead{2T}  & \colhead{1T + IC} }
\startdata
\\[-0.95em]
\textit{kT$_{1}$} (keV) & 9.13 $\pm{0.12}$ & 10.60$^{+0.79}_{-0.44}$  & 7.76$^{+0.26}_{-0.25}$ \\  
\\[-0.5em]
\textit{Z$_{1}$} ({\it Z$_{\odot}$}) & 0.198$\pm{0.026}$  & 0.280$^{+0.034}_{-0.032}$ & 0.282$^{+0.036}_{-0.034}$ \\ 
\\[-0.5em]
\textit{z$_{1}$} & 0.222 & 0.222 & 0.219 \\
\\[-0.5em]
\textit{norm$_{1}$} (10$^{-2}$) & 1.369 $\pm{0.018}$ & 1.139$^{+0.060}_{-0.073}$ & 1.045$^{+0.049}_{-0.046}$ \\
\\[-0.5em]
\textit{kT$_{2}$} & \nodata & 2.16$^{+1.14}_{-0.52}$ & \nodata\\  
\\[-0.5em]
\textit{norm$_{2}$} (10$^{-2}$) & \nodata & 0.728$^{+0.312}_{-0.156}$  & \nodata \\
\\[-0.5em]
$\Gamma$  & \nodata  & \nodata  & 2.0 (fixed) \\ 
\\[-0.5em]
$\kappa$ (10$^{-2}$) & \nodata &\nodata & 0.121$^{+0.016}_{-0.018}$  \\
\\[-0.5em]
{\it C / $\nu$} & 1679.29/1686  & 1638.04/1684  &1640.99/1685  \\ 
\\[-0.5em]
\enddata
\end{deluxetable}

\begin{deluxetable}{lccc}
\tabletypesize{\scriptsize}
\tablewidth{0pt} 
\tablecaption{Same as Table~\ref{tab:globalfit} but for joint \nustar~(3.0~-~20.0 keV) and \chandra~(0.8~-~7.0 keV) joint global spectral fit. All models are convolved with {\tt TBabs}. During the error calculations, redshifts (\textit{z}) are fixed to the best-fit values. Subscripts ``N" refer to \nustar, and ``C" refer to \chandra~parameters.
\label{tab:globalfitNuCh}}
\tablehead{\\[-0.95em]
\colhead{}& \colhead{1T} &\colhead{2T}  & \colhead{1T + IC}}
\startdata
\\[-0.5em]
\textit{N$_{H}$} (10$^{21}$~cm$^{-2}$)& 9.17 $\pm{0.25}$ &  9.80$\pm{0.27}$  & 10.06$\pm{0.33}$\\
\\[-0.5em]
\textit{kT$_{1}$} (keV) & 9.20$^{+0.11}_{-0.12}$ & 21.56$^{+2.71}_{-2.90}$  & 7.99$^{+0.25}_{-0.24}$ \\  
\\[-0.5em]
\textit{Z$_{1N}$} ({\it Z$_{\odot}$}) & 0.202 $\pm{0.026}$  & 0.214 $\pm{0.024}$ & 0.273$^{+0.034}_{-0.033}$ \\
\\[-0.5em]
\textit{Z$_{1C}$} ({\it Z$_{\odot}$}) & 0.550$^{+0.110}_{-0.105}$  & 0.572$^{+0.102}_{-0.097}$ & 0.750$^{+0.135}_{-0.128}$ \\
\\[-0.5em]
\textit{z$_{1N}$} & 0.228 & 0.219 & 0.221 \\
\\[-0.5em]
\textit{z$_{1C}$} & 0.188 & 0.185 & 0.186 \\
\\[-0.5em]
\textit{norm$_{1}$} (10$^{-2}$) & 1.376$\pm{0.017}$ & 0.364$^{+0.095}_{-0.062}$ & 1.081$^{+0.048}_{-0.047}$ \\
\\[-0.5em]
\textit{kT$_{2}$} & \nodata & 6.36$^{+0.40}_{-0.38}$ & \nodata\\  
\\[-0.5em]
\textit{norm$_{2}$} (10$^{-2}$) & \nodata & 1.133$^{+0.061}_{-0.096}$  & \nodata  \\
\\[-0.5em]
$\Gamma$  & \nodata  & \nodata  & 2.0 (fixed) \\ 
\\[-0.5em]
$\kappa$ (10$^{-2}$) & \nodata &\nodata & 0.107$^{+0.018}_{-0.016}$  \\
\\[-0.5em]
{\it C / $\nu$} & 2104.29/2107  & 2067.33/2105  &2071.99/2106   \\ 
\\[-0.5em]
\enddata
\end{deluxetable}

The spectral fit results of the parameters are presented in Table~\ref{tab:globalfit}, and  Table~\ref{tab:globalfitNuCh}, for \nustar, and for joint \nustar~and \chandra, respectively. To study the non-isothermal gas expected from merging clusters of galaxies, as well as possible non-thermal emission due to inverse Compton (IC) scattering, we applied two more models to the spectra following the work of \citet{rojas21} and \citet{wik14}. 

We first added another {\tt apec} component to the 1T model to describe a secondary temperature structure, and fitted the spectra with {\tt constant} $\times$ ({\tt apec} + {\tt apec}), namely 2T model. We then used another model 1T+IC : {\tt constant} $\times$ ({\tt apec} + {\tt powerlaw}), accounting for the thermal emission and a possible IC emission, where the IC component is represented with a power-law distribution. We fixed the photon index to $\Gamma$~=~2 of the {\tt powerlaw} obtained from the spectral index found by \citet{hoang21}.

The addition of a second {\tt apec} and a {\tt powerlaw} to the original single temperature model for \nustar~ spectrum, resulted in an improvement of {\it $\Delta$C/$\Delta\nu$} = 41.25/2 and {\it $\Delta$C/$\Delta\nu$} = 38.30/1 for 2T and 1T + IC model, respectively, with respect to single temperature fit. For the joint \nustar~and \chandra~analysis, the fit also improved by an additional {\tt apec} or a {\tt powerlaw} component, where we found {\it $\Delta$C/$\Delta\nu$} = 36.96/2 and {\it $\Delta$C/$\Delta\nu$} = 32.30/1 for 2T and 1T + IC, respectively. Both \nustar, and joint \nustar~and \chandra~analysis show that the central \textit{r} = 5.2\arcmin emission is best described by the 2T model.

We also calculated the luminosity of the cluster from the \nustar~spectra within this region (r = 312$\arcsec$ $\simeq$ 0.6 R$_{500}$ adopting R$_{500}$ = 500$\arcsec$ \citep{reiprich13}). We find the X-ray bolometric (0.01-100 keV) luminosity to be 
{\it L$_{X,bol}$} = 1.390~$\times$~10$^{45}$ erg s$^{-1}$, and within 0.5-2.0 keV we find {\it L$_{X,0.5-2.0keV}$} = 2.716~$\times$~10$^{43}$ erg s$^{-1}$ using ${\tt XSPEC}$ convolution model ${\tt clumin}$, for the 1T model. 

\nustar~gain issue is usually accommodated by the use of {\tt xspec} {\tt gain} command. The gain over the years shows an offset value around -0.1, but is still a work under construction (Rojas Bolivar et al., submitted). In the global spectral analysis using 1T model, we found that the gain is -0.15. When we use the {\tt gain} command, the temperature becomes \textit{kT}=9.15$^{+0.17}_{-0.12}$ keV with 1684.29/1685 d.o.f., and the temperature is \textit{kT}=9.13$\pm{0.12}$ keV with 1679.29/1686 d.o.f., when the redshift is freed to vary. Using the {\tt gain} command instead of freeing the redshift changes the temperature by only 0.3\%, but does not necessarily provide a better model. Due to the complexity of applying the {\tt gain} command further in our analyses, we chose to free the redshift to account for the gain issue for the rest of this work.

In addition, here we share the luminosities that account for the evolution of the {\it L$_{X}$-T} scaling relations, using E(z)=1.09, where E(z)=$\sqrt{\Omega_{M}(1+z)^3+(1-{\Omega_{M}}-\Omega_{\Lambda})(1+z)^2+\Omega_{\Lambda}}$ \citep{giles16}. Diving the obtained luminosities by E(z), the luminosity values become {\it L$_{X,bol}$} = 1.275~$\times$~10$^{45}$ erg s$^{-1}$ and {\it L$_{X,0.5-2.0keV}$} = 2.492~$\times$~10$^{43}$ erg s$^{-1}$.

The resulting spectra are shown in Section~\ref{sec:globalfit}.

\subsection{Surface brightness discontinuities} \label{sec:SB}

To study the correspondences of northern and southern \chandra~SB edges (Fig.~\ref{fig:nuphoton}) with \nustar, we used Gaussian Gradient Magnitude (GGM) filter\footnote{\url{https://github.com/jeremysanders/ggm}} \citep{sanders16}. Assuming Gaussian derivatives with a width of $\sigma$, the GGM filter calculates the gradient of an image. To create the GGM filtered image, the image itself is convolved by the gradient of a 1D Gaussian function for two axes, then these two resulting images are combined for the 2D gradient image.
The width, $\sigma$, is varied to capture gradients on different scales, i.e., small $\sigma$ values are used at the central regions where there are many counts, and high values are better capturing the gradient at cluster outskirts with low counts.

\begin{figure}
\centering
\includegraphics[width=86mm]{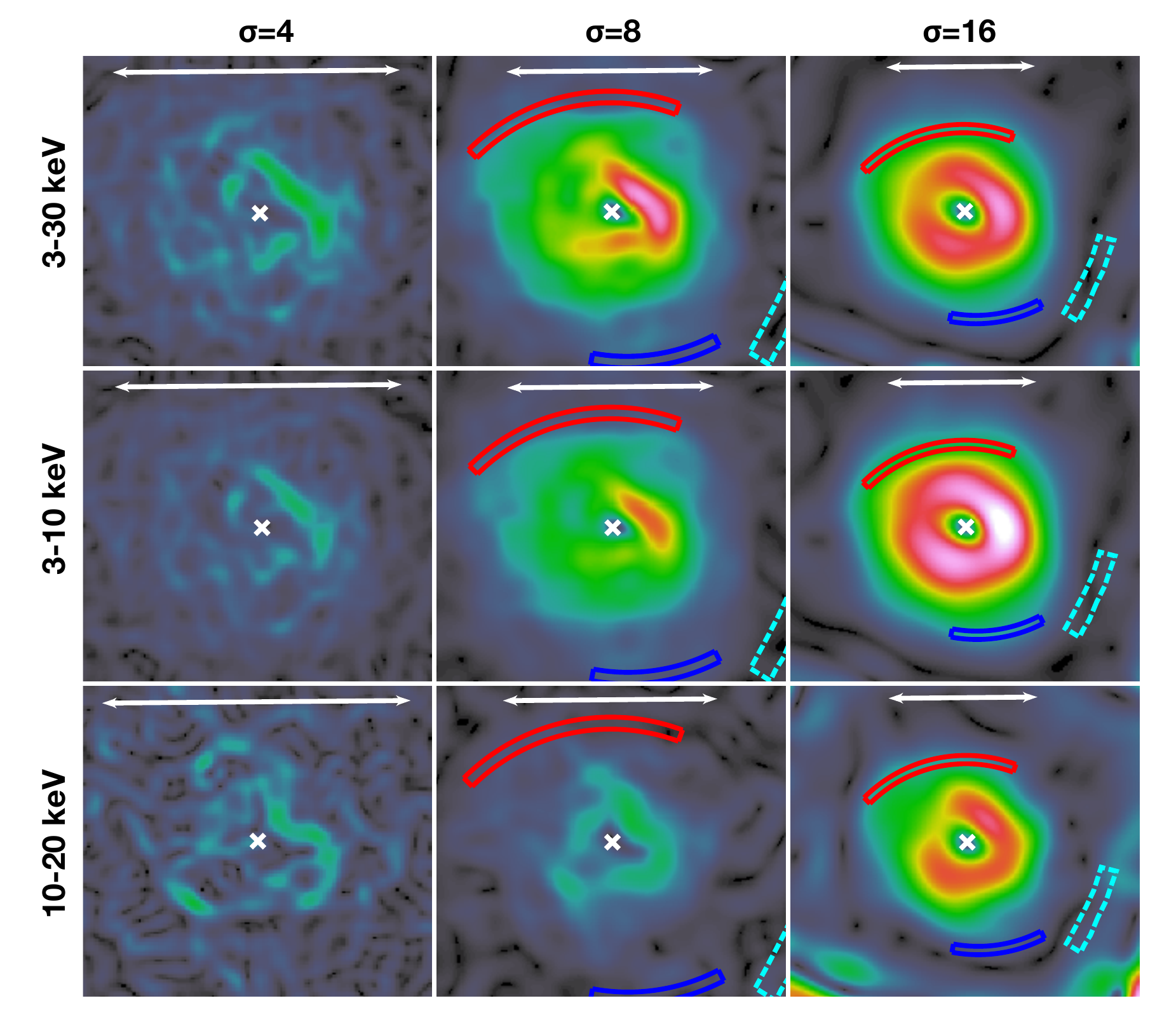}
\caption{\nustar~GGM filtered background subtracted, exposure corrected images in the 3.0~-~30.0 keV ({\it top panels}), 3.0~-~10.0 keV ({\it middle horizontal panels}), and 10.0~-~20.0 keV ({\it lower panels}) energy bands. White cross correspond to the emission peak, red (blue) arc corresponds to the northern (southern) SB edge, and dashed turquoise region represents the channel. The white arrow in all images correspond to 1 Mpc. \label{fig:GGM}}
\end{figure}

\begin{figure}
\centering
\includegraphics[width=82mm]{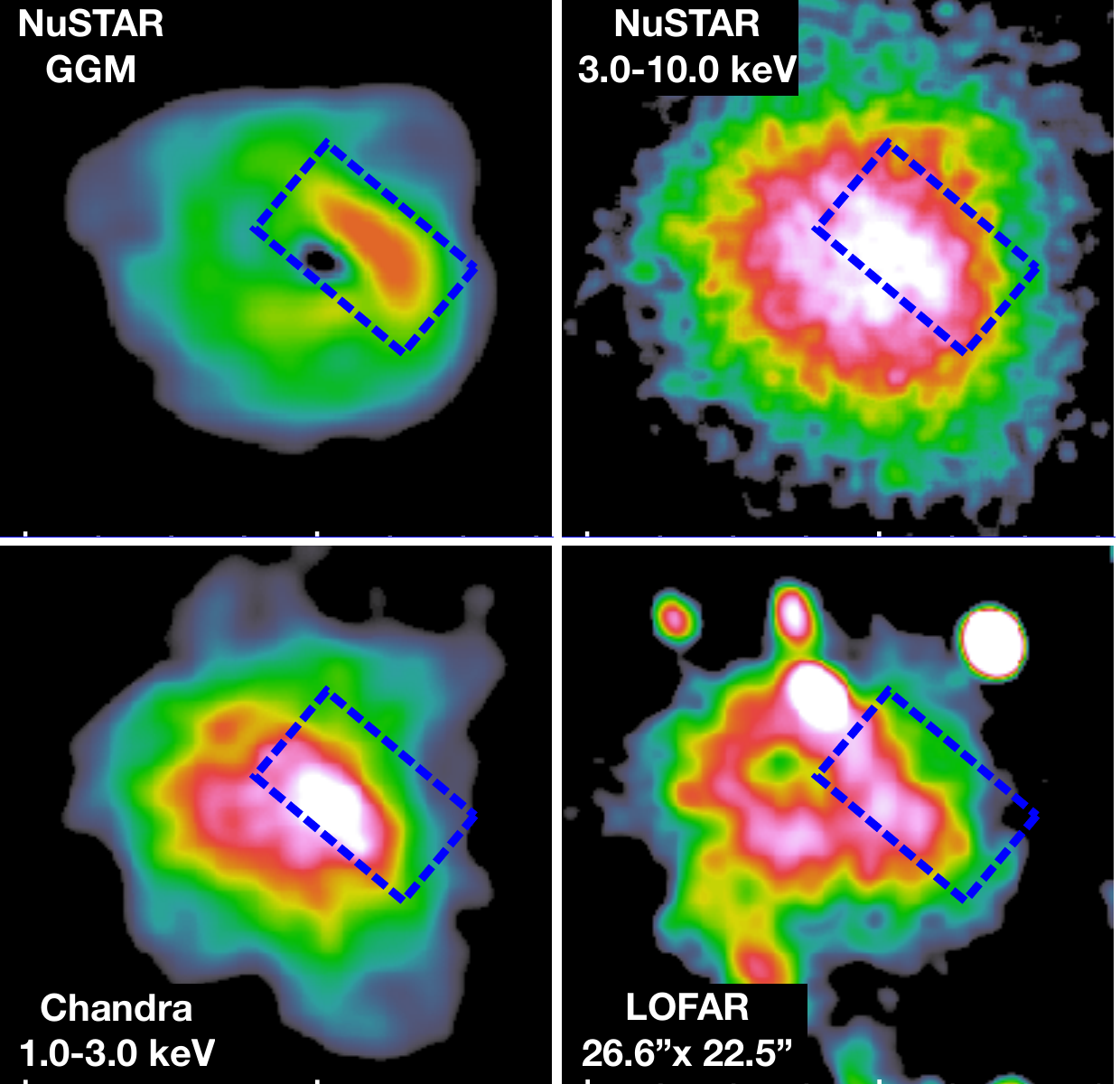}
\caption{$\sigma$ = 8, 3.0~-~10.0 keV \nustar~GGM ({\it upper left }), \nustar~ 3.0~-~10.0 keV photon ({\it upper right}), \chandra~1.0~-~3.0 keV photon ({\it lower left}), and LOFAR 26.6$\arcsec$ $\times$ 22.5$\arcsec$ resolution ({\it lower right}) images. Dashed blue box corresponds to the same regions in all images \label{fig:inneredge}.}
\end{figure}

Since \nustar~PSF is much larger than the \chandra~PSF for which the GGM filter was intended for, the filters with low levels of sigma ($\sigma$ = 1, 2) were mainly dominated by the noise. Therefore, we present the GGM filtered to the background subtracted, exposure corrected images with using $\sigma$ = 4, 8, and 16 detector pixels, as shown in Fig.~\ref{fig:GGM}. 

Due to the large PSF of \nustar, the northern and southern SB edges are not unambiguously visible from the GGM results. We note that the SB edge and the channel locations obtained from \citet{zhang20} are overlaid on the GGM (Fig.~\ref{fig:GGM}) merely to guide the eye and are not indicated by GGM analysis. The use of the GGM in this context is to reveal the strong SB feature much closer to the cluster center than the \chandra~detected northern and southern SB edges.

GGM point to a sharp gradient best discerned in the panels presenting $\sigma$ = 8, 3.0~-~10.0 and 3.0~-~30.0 keV band, and  $\sigma$ = 16, 3.0~-~10.0 keV band, suggesting a SB edge. This edge can also be seen in \nustar, \chandra, and LOFAR images. For a better visualization of this edge, we present the \nustar~GGM, \nustar~photon, \chandra~photon, and LOFAR images together in Fig.~\ref{fig:inneredge}.

\subsection{Temperature map} \label{sec:TempMap}

To study the temperature structure of the \nustar\ FOV, we created a temperature map. We extracted count images from a 12$\arcmin$ $\times$ 12$\arcmin$ central box region of the cluster in the 3.0~-~5.0, 6.0~-~10.0, and 10.0~-~20.0 keV bands where the point sources detected by \chandra\ were excluded. We avoided the 5.0~-~6.0 keV band due to the existence of FeK complex. The background images and exposure maps were extracted for the same energy bands with {\tt nuskybgd} and {\tt nuproducts}, respectively. 

\begin{figure}[h!]
\includegraphics[width=84mm]{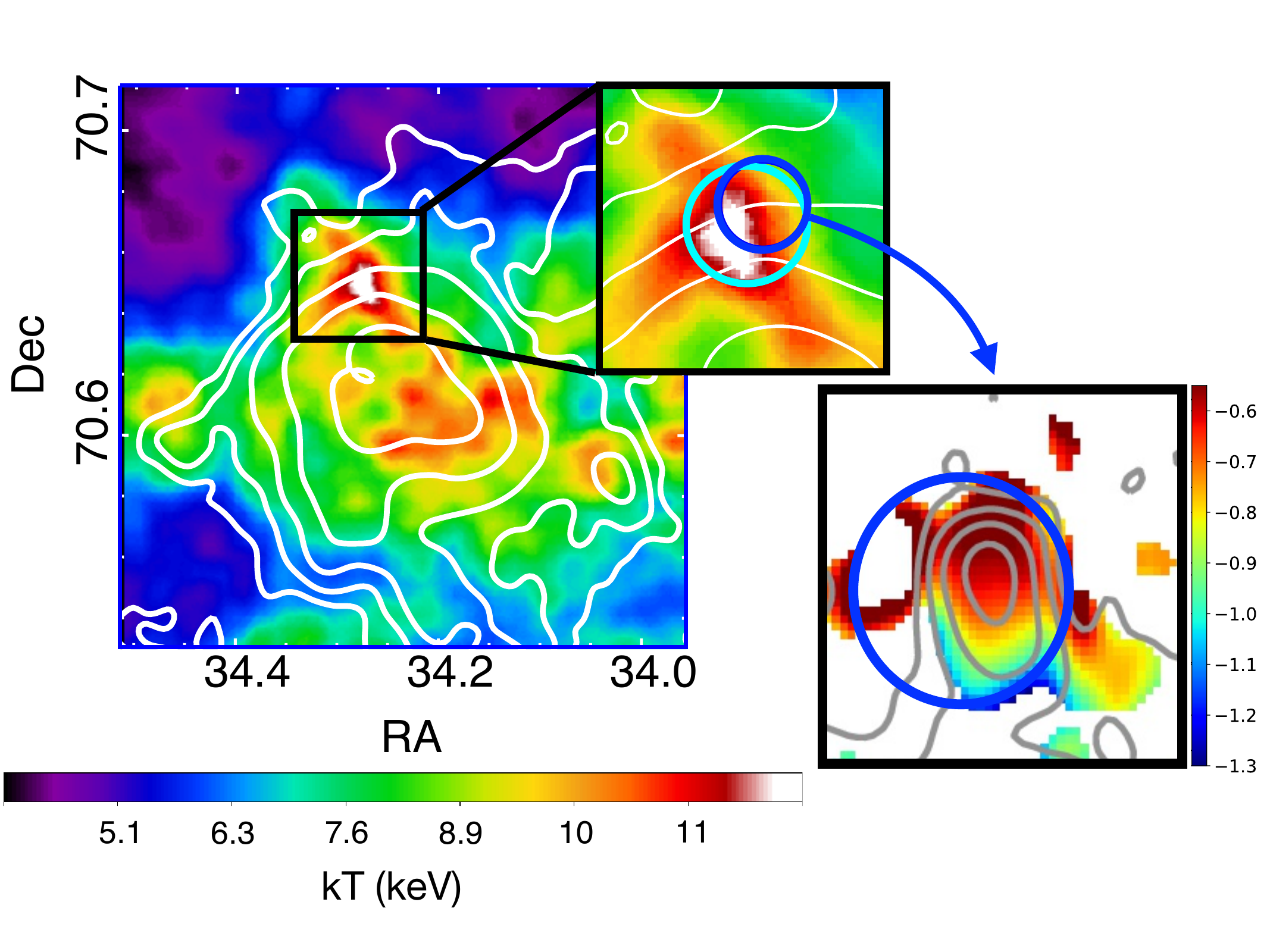}
\caption{\nustar~temperature map with radio contours from LOFAR (white, at $\sigma_{rms}\times$ [3, 6, 12, 24, 48], where $\sigma_{rms}$ = 330~$\mu$Jy/beam [46$\arcsec\times$45$\arcsec$]) are overlaid. In the zoomed in panel (black box, upper right corner), the ``hot spot" is indicated by turquoise circle, and the radio source is shown with blue circle. In the lower right corner, the radio galaxy spectral map between 141 MHz (LOFAR) and 1.5 GHz (VLA) at 16$\arcsec$ resolution is shown \citep[][Fig. 6]{hoang21}. \label{fig:TempMap}}
\end{figure}

The background subtracted, exposure corrected images were then used for the spectral extraction and predefined pixels were fitted with a single temperature {\tt apec} model, during which the abundance and redshift were fixed to the values obtained from the global fit. The technique is described in detail by \citet{markevitch00}. The resulting temperature map is presented in Fig.~\ref{fig:TempMap}. We emphasize that this temperature map is a very basic analysis tool to search for any strong features in the ICM as in the case of the GGM analysis. We present more quantitative, accurate and precise temperature results further in the work.

In the temperature map, we encountered a ``hot spot"\footnote{``Hot spot" refers to the feature of unknown origin seen in the temperature map, while without quotations, we refer to a thermal plasma emission.} that has a higher temperature than the cluster core. In the vicinity ($\sim$ 0.5$\arcmin$) of the ``hot spot" (Fig.~\ref{fig:TempMap}, turquoise circle), there is a radio source GB3 0212+704\footnote{\url{https://vizier.cds.unistra.fr/viz-bin/VizieR-5?-ref=VIZ620b3ce430f570&-out.add=.&-source=VIII/53/gb1&recno=3648}} \citep{maslowski72} (Fig.~\ref{fig:TempMap}, blue circle) and LOFAR high resolution study shows a tail morphology. The spectral index map between LOFAR 145 MHz and VLA 1.5 GHz also shows the steepening of the spectrum behind the tail. The tail radio galaxy has flux density of S (141 MHz) = 12.1$\pm$1.2 mJy, S (1.4 GHz)= 2.9 +/- 0.2 mJy, S (1.5 GHz)= 2.7 +/- 0.2 mJy. The average spectral index is $\alpha$= -0.62 $\pm$ 0.01. 

\begin{deluxetable}{lccc}
\tabletypesize{\scriptsize}
\tablewidth{0pt} 
\tablecaption{Spectral parameters and 1$\sigma$ uncertainty ranges of \nustar\ spectrum fits in 3.0~-~20.0 band of the ``hot spot" scenarios.
\label{tab:hotspotscenario}}
\tablehead{\\[-0.95em]
\colhead{}& \colhead{Hot spot} &\colhead{Obscured AGN}  & \colhead{IC}}
\startdata
\\[-0.5em]
\textit{N$_{H}$} (10$^{22}$~cm$^{-2}$)& \nodata &  5.11 $^{+3.63}_{-3.52}$  & \nodata\\
\\[-0.5em]
\textit{kT} (keV) & 12.10$^{+2.01}_{-1.29}$ & \nodata  & \nodata \\  
\\[-0.5em]
\textit{Z} ({\it Z$_{\odot}$}) & 0.612 $^{+0.388}_{-0.312}$  & \nodata & \nodata \\
\\[-0.5em]
\textit{z} & 0.249 $^{+0.046}_{-0.033}$ & \nodata & \nodata \\
\\[-0.5em]
\textit{norm} (10$^{-4}$) & 1.497 $^{+0.187}_{-0.158}$ & \nodata & \nodata \\
$\Gamma$  & \nodata  & 2.34 $^{+0.18}_{-0.17}$ & 2.13 $^{+0.09}_{-0.08}$ \\ 
\\[-0.5em]
$\kappa$ (10$^{-4}$) & \nodata & 1.100$^{+0.542}_{-0.356}$ & 0.657$^{+0.120}_{-0.102}$  \\
\\[-0.5em]
{\it C / $\nu$} & 381.91/364  & 382.37/365  &384.50/366   \\ 
\\[-0.5em]
\enddata
\end{deluxetable}

To study this feature in detail with \nustar, we extracted a circular region with r = 0.6$\arcmin$ ($\sim$112 kpc) enclosing the most dominant region of this ``hot spot", centered at 2:17:04.08 (RA) and +70:39:06 (Dec). The temperature map indicates that, if of thermal origin, the temperature of the plasma is $\geq$~12 keV. This feature may be a part of a more extended emission, yet it is difficult define this extension.

To explain its nature, we put forward three scenarios. Scenario 1 suggests that the feature is the downstream shock region evidenced by its location on the northern \chandra~SB edge. Scenario 2 is an emission from a heavily obscured AGN (since \chandra~doesn not detect it) that may be due to a strong chaotic cold accretion (CCA) rain that feeds the central supermassive black hole \citep{gaspari13a,gaspari20}. Scenario 3 is a localized IC emission where the electrons in the nearby radio source is upscattered to X-ray regime by a possible shock front.

We first fit the \nustar~spectra with a single temperature {\tt apec} model (Scenario 1). The temperature obtained from this fit was {\it kT} = 12.1$^{+2.0}_{-1.3}$ keV with {\it C/$\nu$} = 381.91/364. To test Scenario 2, we used a {\tt powerlaw} model convolved with a {\tt TBabs} model to account for a heavily obscured AGN. The resulting hydrogen column density value is N$_H$ = 5.11$\times$10$^{22}$~cm$^{-2}$, where the photon index for the hypothesized AGN emission is $\Gamma$ = 2.34$^{+0.18}_{-0.17}$. The statistics of this fit is {\it C/$\nu$} = 382.37/365. And for Scenario 3, we used an unabsorbed {\tt powerlaw} model to describe possible IC emission. The resulting powerlaw slope is $\Gamma$ = 2.13$^{+0.09}_{-0.08}$. The statistics of this fit is {\it C/$\nu$} = 384.50/366. The corresponding values and spectra of the fits from these scenarios are presented in Table~\ref{tab:hotspotscenario} and in Section~\ref{sec:hotspot}, respectively. Statistics show that these three scenarios are equally likely.

Lastly, to rule out a possible multi-temperature structure, we also fitted a two temperature model to the spectrum. The high temperature component rose up to {\it kT} $\sim$ 28.8 keV, where the low component was {\it kT} $\sim$ 7.41. In this fit, both temperature parameters were unconstrained. 

\begin{figure*}
\centering
\includegraphics[width=134mm]{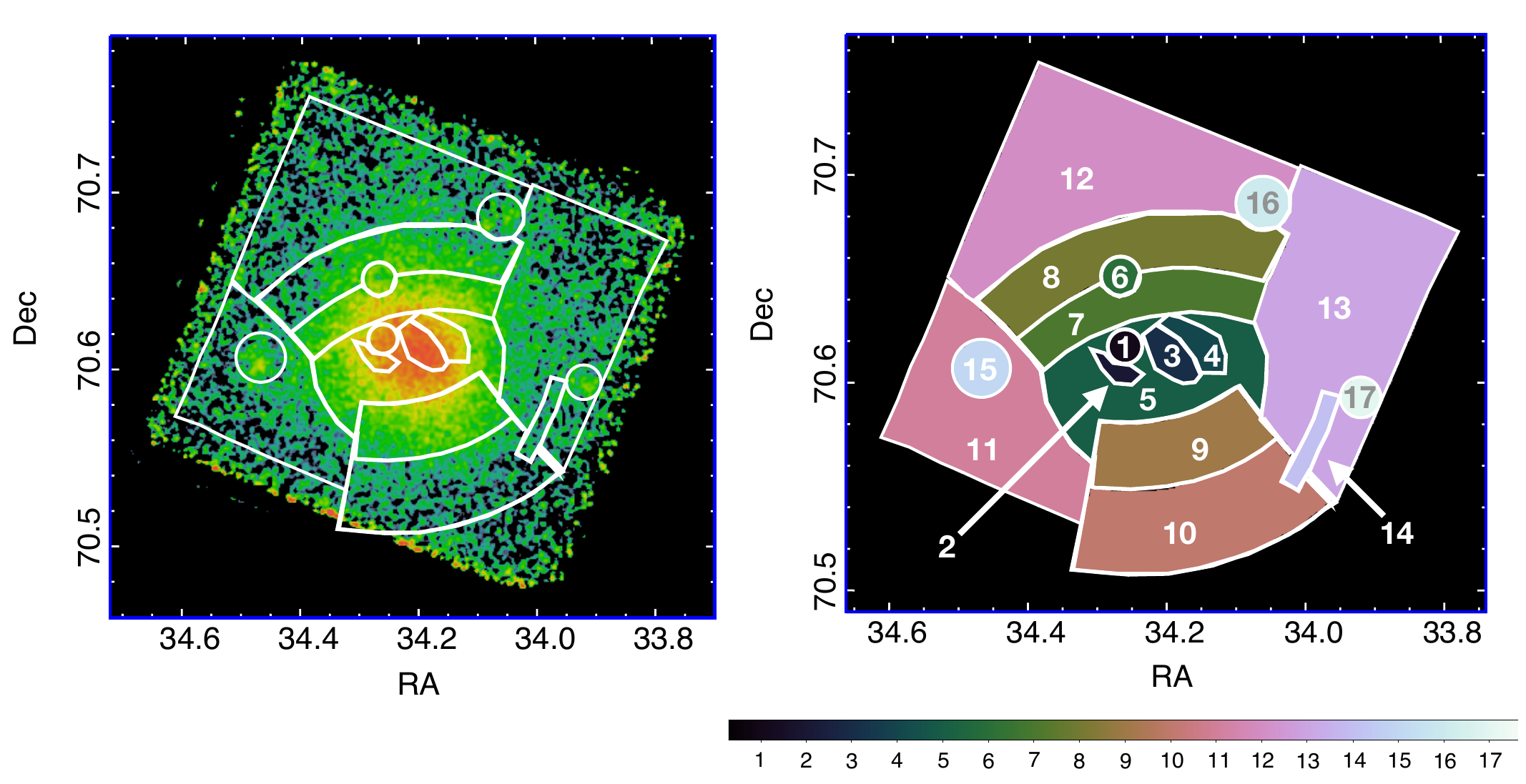}
\caption{Regions selected for the cross-talk assessment. \label{fig:crossarfmask}}
\end{figure*}

\subsection{Crosstalk analysis} \label{sec:Crosstalk}

Even single bright sources in the \nustar~FOV causes the scatter photons due to the large ($\sim$1$\arcmin$ Half Power Diameter, $\sim$18$\arcsec$ Full Width at Half Maximum), slightly energy-dependent point spread function (PSF). This results in a cross-contamination, namely crosstalk, of multiple emission sources in the regions of interest, although the emission may not originate from the selected region. {\tt nuproducts} produces ARFs for point or diffuse sources inside the user defined extraction regions. However, it does not account for the ARFs for other sources whose emission originates outside these extraction regions that contaminate the spectra of those regions. These ARFs will be referred to as cross-ARFs.

Software composed of a set of routines created to account for this contamination is {\tt nucrossarf}\footnote{\url{https://github.com/danielrwik/nucrossarf}}.

In order to study the temperature jumps and prominent structures in the FOV more accurately, we selected regions of interest in the \nustar\ data and used {\tt nucrossarf} to account for the crosstalk. The selected regions are overlaid on the \nustar\ image on the left panel, and the corresponding region values are presented in the right panel of Fig.~\ref{fig:crossarfmask}. 

\begin{deluxetable*}{cccccccccc}
\tabletypesize{\scriptsize}
\tablewidth{0pt} 
\tablecaption{Crosstalk analysis spectral fit results where Region 6 is treated as a hot spot, obscured AGN and inverse Compton emission. In the obscured AGN scenario, the value \textit{N$_{H}$} is kept fixed at  5.11 $\times$ 10$^{22}$~cm$^{-2}$. ${\tt apec}$ normalization ({\it norm}) is given in $\frac{10^{-14}}{4\pi \left [ D_A(1+z) \right ]^{2}}\int n_{e}n_{H}dV$ (10$^{-3}$~cm$^{-5}$) where ${\tt powerlaw}$ normalization ($\kappa$) is {\it photons~keV$^{-1}$~cm$^{-2}$~s$^{-1}$} at 1 keV (10$^{-3}$).
\label{tab:nuprocrossfit}}
\tablehead{\\[-0.95em]
\colhead{}& \multicolumn{4}{c}{$kt$ (keV) or $\Gamma$} & \multicolumn{4}{c}{{\it norm} or $\kappa$} & \colhead{} \\[-0.95em]
\colhead{Region}& \colhead{{\tt nuproducts}} & \multicolumn{3}{c}{{\tt nucrossarf}} & \colhead{{\tt nuproducts}} & \multicolumn{3}{c}{{\tt nucrossarf}}&\colhead{Notes} \\[-0.95em]
\colhead{Number}& \colhead{}& \colhead{Hot spot} &\colhead{AGN}  & \colhead{IC} &\colhead{}& \colhead{Hot spot} &\colhead{AGN}  & \colhead{IC} & \colhead{}}
\startdata
\\[-0.95em]
1 & 10.18$^{+0.68}_{-0.33}$& 9.90$^{+1.90}_{-1.38}$& 9.89$^{+1.90}_{-1.38}$& 9.90$^{+1.90}_{-1.38}$
& 0.356$^{+0.023}_{-0.021}$  & 0.459$^{+0.056}_{-0.051}$ & 0.460$^{+0.060}_{-0.051}$&0.459$^{+0.056}_{-0.051}$&\\
\\[-0.5em]
2& 9.47$^{+0.62}_{-0.73}$&7.47$^{+2.44}_{-1.88}$&7.48$^{+2.46}_{-1.87}$& 7.43$^{+2.45}_{-1.85}$
& 0.235$^{+0.018}_{-0.019}$& 0.217$^{+0.061}_{-0.042}$ &0.218$^{+0.060}_{-0.042}$&0.219$^{+0.061}_{-0.042}$& \\
\\[-0.5em]
3& 11.18$^{+0.43}_{-0.20}$&12.53$^{+0.94}_{-0.83}$&12.55$^{+0.95}_{-0.82}$& 12.51$^{+0.94}_{-0.83}$
& 0.996$^{+0.038}_{-0.034}$ & 1.406$^{+0.060}_{-0.048}$ &1.409$^{+0.060}_{-0.048}$& 1.407$^{+0.061}_{-0.049}$& C$_{in}$\\
\\[-0.5em]
4& 10.61$^{+0.49}_{-0.46}$&10.29$^{+0.93}_{-0.71}$& 10.25$^{+0.91}_{-0.71}$& 10.31$^{+0.92}_{-0.71}$
& 0.573$^{+0.030}_{-0.028}$& 0.800$^{+0.041}_{-0.042}$& 0.801$\pm{0.041}$&0.799$^{+0.041}_{-0.042}$& C$_{out}$\\
\\[-0.5em]
5& 9.92$^{+0.16}_{-0.22}$&9.37$^{+0.42}_{-0.44}$&9.35$^{+0.42}_{-0.45}$ & 9.40$^{+0.42}_{-0.44}$
& 3.452$^{+0.089}_{-0.063}$& 3.626$^{+0.134}_{-0.123}$ &3.635$^{+0.139}_{-0.124}$&3.617$^{+0.134}_{-0.121}$& \\
\\[-0.5em]
6& 12.02$^{+1.76}_{-1.31}$&21.02 \tablenotemark{a} &\nodata & \nodata 
& 0.149$^{+0.017}_{-0.016}$& 0.108$\pm{0.012}$ &\nodata &\nodata & Hot spot\\
\\[-0.5em]
6& 2.34$^{+0.18}_{-0.17}$&\nodata &2.34\tablenotemark{b} & \nodata 
& 0.110$^{+0.054}_{-0.036}$& \nodata  & 0.054$\pm{0.007}$&\nodata & AGN \\
\\[-0.5em]
6& 2.13$^{+0.09}_{-0.08}$&\nodata &\nodata & 1.38$\pm{0.26}$
& 0.066$^{+0.012}_{-0.010}$& \nodata  &\nodata &0.014$^{+0.010}_{-0.006}$& IC\\
\\[-0.5em]
7& 9.29$^{+0.27}_{-0.16}$&8.31$^{+0.43}_{-0.41}$&8.57$^{+0.45}_{-0.41}$& 8.12$^{+0.44}_{-0.40}$
& 1.862$^{+0.051}_{-0.027}$ & 2.028$^{+0.088}_{-0.082}$ &2.014$^{+0.084}_{-0.081}$&2.067$^{+0.094}_{-0.089}$& N$_{in}$\\
\\[-0.5em]
8& 7.51$^{+0.30}_{-0.53}$&6.70$^{+0.52}_{-0.47}$&6.98$^{+0.54}_{-0.49}$& 6.56$^{+0.51}_{-0.47}$
& 1.449$^{+0.055}_{-0.114}$ & 1.342$^{+0.098}_{-0.090}$ &1.319$^{+0.093}_{-0.086}$&1.374$^{+0.105}_{-0.095}$& N$_{out}$\\
\\[-0.5em]
9& 8.98$^{+0.39}_{-0.31}$&8.19$^{+0.45}_{-0.44}$&8.20$^{+0.45}_{-0.43}$& 8.18$^{+0.45}_{-0.43}$
& 1.527$^{+0.061}_{-0.053}$ & 1.575$^{+0.072}_{-0.064}$ &1.578$^{+0.072}_{-0.064}$&1.577$^{+0.072}_{-0.065}$& S$_{in}$ \\
\\[-0.5em]
10& 8.05$^{+0.72}_{-0.59}$&6.02$^{+0.97}_{-0.72}$&6.02$^{+0.97}_{-0.72}$& 6.02$^{+0.98}_{-0.72}$
& 0.819$^{+0.055}_{-0.054}$ & 0.645$^{+0.090}_{-0.082}$ &0.646$^{+0.090}_{-0.082}$&0.645$^{+0.090}_{-0.082}$& S$_{out}$\\
\\[-0.5em]
11& 7.13$^{+0.54}_{-0.42}$&4.46$^{+0.57}_{-0.62}$&4.44$^{+0.56}_{-0.63}$& 4.46$^{+0.57}_{-0.61}$
& 1.195$^{+0.075}_{-0.071}$ & 0.956$^{+0.177}_{-0.112}$ &0.960$^{+0.181}_{-0.120}$&0.953$^{+0.175}_{-0.119}$& \\
\\[-0.5em]
12& 4.72$^{+0.29}_{-0.37}$&3.16$^{+0.22}_{-0.20}$&3.15$^{+0.22}_{-0.20}$& 3.18$^{+0.23}_{-0.21}$
& 2.921$^{+0.226}_{-0.206}$ & 3.765$^{+0.427}_{-0.380}$ &3.803$^{+0.426}_{-0.385}$&3.740$^{+0.427}_{-0.379}$& \\
\\[-0.5em]
13& 9.97$^{+0.82}_{-0.52}$&7.40$^{+0.64}_{-0.60}$&7.37$^{+0.63}_{-0.61}$& 7.42$^{+0.65}_{-0.60}$
& 1.819$^{+0.104}_{-0.077}$ & 1.667$^{+0.112}_{-0.107}$&1.672$^{+0.123}_{-0.107}$&1.664$^{+0.112}_{-0.107}$& \\
\\[-0.5em]
14& 8.25$^{+2.26}_{-1.94}$&2.17$^{+.....}_{-1.29}$\tablenotemark{c}&2.20$^{+.....}_{-1.31}$\tablenotemark{c}& 2.17$^{+.....}_{-1.29}$\tablenotemark{c}
& 0.052$^{+0.016}_{-0.013}$ & 0.140$^{+......}_{-0.130}$\tablenotemark{d} &0.138$^{+......}_{-0.128}$\tablenotemark{d}&0.141$^{+......}_{-0.131}$\tablenotemark{d}& Channel\\
\\[-0.5em]
15& 2.17$\pm{0.13}$& 2.17\tablenotemark{b} &2.17\tablenotemark{b} &2.17\tablenotemark{b}
& 0.076$^{+0.021}_{-0.017}$ & 0.040$\pm{0.006}$ &0.040$\pm{0.006}$ & 0.040$\pm{0.006}$ & PS\\
\\[-0.5em]
16& 1.76$^{+0.14}_{-0.13}$& 1.76\tablenotemark{b} &1.76\tablenotemark{b} &1.76\tablenotemark{b}
& 0.041$^{+0.012}_{-0.010}$& 0.019$\pm{0.003}$ &0.019$\pm{0.003}$&0.019$\pm{0.003}$& PS \\
\\[-0.5em]
17& 2.79$\pm{0.17}$& 2.79\tablenotemark{b} &2.79\tablenotemark{b} &2.79\tablenotemark{b}
& 0.183$^{+0.062}_{-0.046}$& 0.118$^{+0.017}_{-0.016}$ &0.118$\pm{0.016}$&0.118$\pm{0.016}$& PS\\
\\[-0.95em]
\enddata
\tablenotetext{a}{Lower limit.}
\tablenotetext{b}{Photon indices are fixed to the {\tt nuproducts} values.}
\tablenotetext{c}{Upper temperature limit hits 9 keV.}
\tablenotetext{d}{Upper norm limit hits 1 $\times$ 10$^{-3}$~cm$^{-5}$.}
\end{deluxetable*}

\begin{deluxetable}{ccccc}
\tabletypesize{\scriptsize}
\tablewidth{0pt} 
\tablecaption{Mach numbers of shock fronts calculated from the temperature jumps obtained from the crosstalk results..
\label{tab:mach}}
\tablehead{\\[-0.95em]
\multicolumn{2}{c}{Regions}& \multicolumn{3}{c}{Mach Number}\\
\colhead{Post-shock}& \colhead{Pre-shock}& \colhead{Hot spot} &\colhead{Obscured AGN}  & \colhead{IC}}
\startdata
\\[-0.5em]
3 & 4 & 1.22 & 1.22 & 1.21 \\  
6&8 & 2.01  & \nodata & \nodata \\
7&8 & 1.24 & 1.23 & 1.24 \\
9&10 & 1.37 & 1.37 & 1.37 \\
\enddata
\end{deluxetable}

The motivation behind selecting regions are based on the recent \chandra~study proposing surface brightness edges, the results of our \nustar~GGM and temperature map results, and LOFAR images. Region 1 corresponds to the radio SB depression as seen in the LOFAR image (Fig.~\ref{fig:lofar}, right panel). Region 2 is a wedge, again evidenced by LOFAR image. We selected these regions to investigate any correspondence of these radio features with the X-ray data. Regions 3 (C$_{in}$) and 4 (C$_{out}$) correspond to the inner and outer side of the central SB edge, respectively, as highlighted in Fig.~\ref{fig:inneredge}. Region 5 represents the central region of the cluster that excludes the emission from Regions 1, 2, 3, and 4. Region 6 is the ``hot spot", which we investigate with the aforementioned three scenarios in Section~\ref{sec:TempMap}. 

Regions 7 (N$_{in}$) and 8 (N$_{out}$) are the inner (downstream) and outer (upstream) part of the northern SB edge proposed by \citet{zhang20}, respectively, and similarly, Regions 9 (S$_{in}$) and 10 (S$_{out}$) are the inner and outer sides of the southern SB, respectively. We note that Regions 7, 8, 9 and 10 are not the exact regions selected by \citet{zhang20} that were used to draw the SB profiles on the corresponding edges in their work. The \nustar~FOV ($13\arcmin \times 13\arcmin$) is considerably smaller than the \chandra~FOV ($16\arcmin \times 16\arcmin$), and the regions they use for the SB profiles exceeds our FOV. We widened the angle of the southern and northern edge arcs to be able to capture as much upstream photons as possible for better statistics.

Regions 11, 12 and 13 indicate the outer cluster emission at different directions that is most likely isothermal. Channel is represented by Region 14. Finally, Regions 15, 16, and 17 are point sources as seen by \chandra~and \nustar.

We extracted spectra from these 17 regions in total using {\tt nuproducts}, and found the best fit model for these regions. In total, we have 13 regions to account for the ICM emission modeled by an {\tt apec}, and 3 point sources (circular regions in Fig.~\ref{fig:crossarfmask}) modeled by a {\tt powerlaw}. And for Region 6, we use three different models as explained in detail in Section~\ref{sec:TempMap}.

During the {\tt nucrossarf} fit, we fitted 17 spectra, and their individual 17 ARFs simultaneously, where we fixed the redshift and abundance values to the individual {\tt nuproducts} spectral fits. For Region 6, we studied the three scenarios. We ran the {\tt nucrossarf} code thrice, first treating Region 6 as a hot spot, then obscured AGN and finally, localized IC emission.
The results from the {\tt nuproducts} and {\tt nucrossarf} spectral analysis are given in Table~\ref{tab:nuprocrossfit}. 

We note that Region 14, where \chandra~detects the possible SB depression (Channel), the upper limits for kT and {\tt apec} normalization were quite large, therefore we constrained the upper limits to 9 keV and 1 $\times$ 10$^{-3}$~cm$^{-5}$ for the temperature and {\tt apec} normalization, respectively. 

For all of the AGN powerlaw models, we fixed the photon index to the {\tt nuproducts} value to prevent an artificial gauge of any excess in the hard band regime by the powerlaw model. Since these regions are treated as point sources and since the regions attributed to them is comparable to the \nustar~PSF, their emission should be localized, meaning the shape of the slope would be conserved.

\begin{figure}
\centering
\includegraphics[width=84mm]{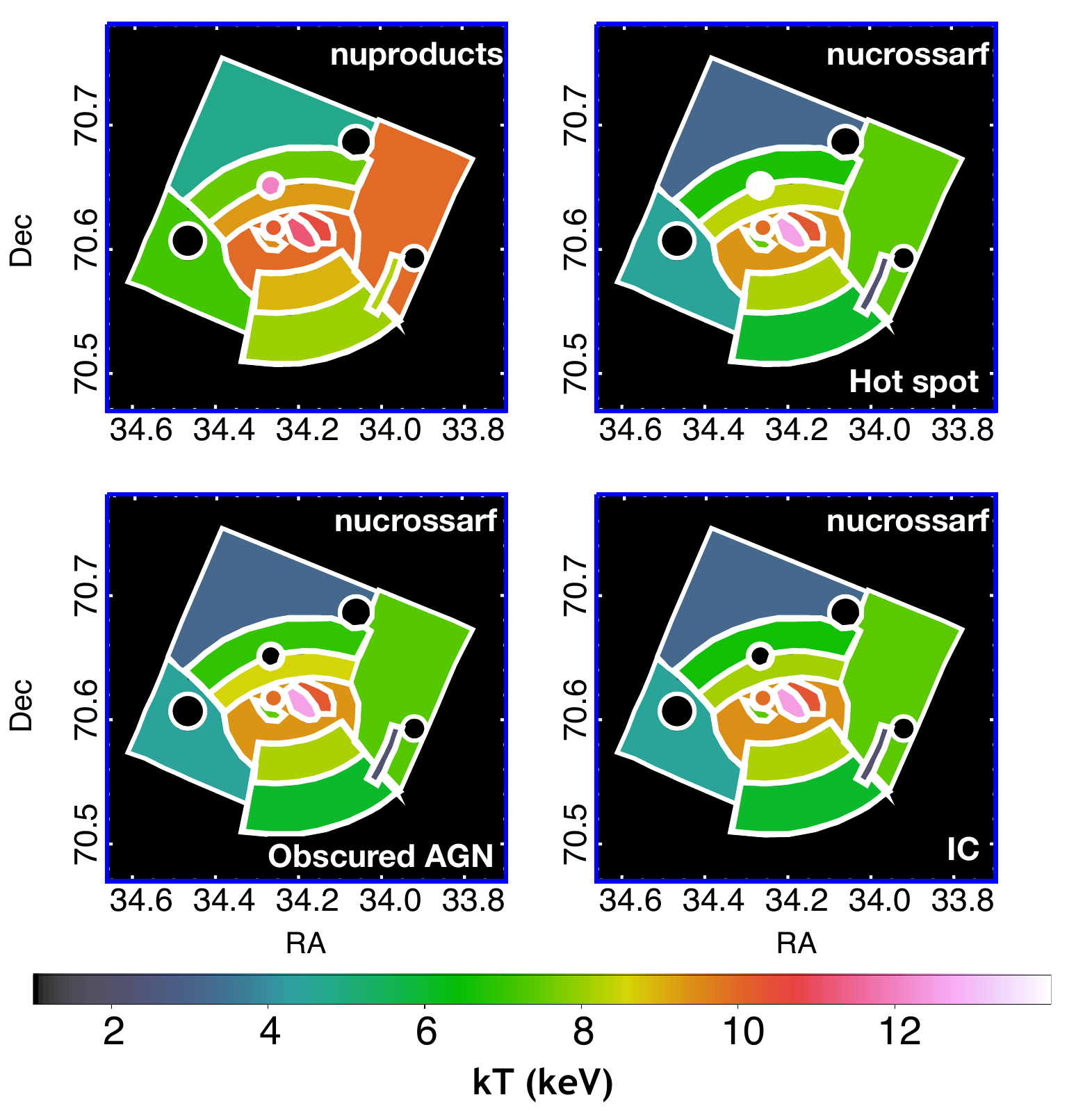}
\caption{Temperature maps created with the results of {\tt nuproducts} and {\tt nucrossarf} analysis. We use color black to fill in the regions where a power-law model is used to describe the data, that is regions 15, 16 and 17 for all maps, and Region 6 for obscured AGN and IC scenarios. \label{fig:crosstalkresults}}
\end{figure}

To better visualise the effect of the inclusion of crosstalk analysis, we present temperature maps created by filling the regions of interest with the obtained temperature values from the fits in Fig.~\ref{fig:crosstalkresults}. This figure showing the temperature results from all scenarios are also a testament to the stability of the crosstalk analysis, as they visualize the almost identical temperature values when we assume difference emission models for Region 6. 

\begin{figure*}
\centering
\includegraphics[width=180mm]{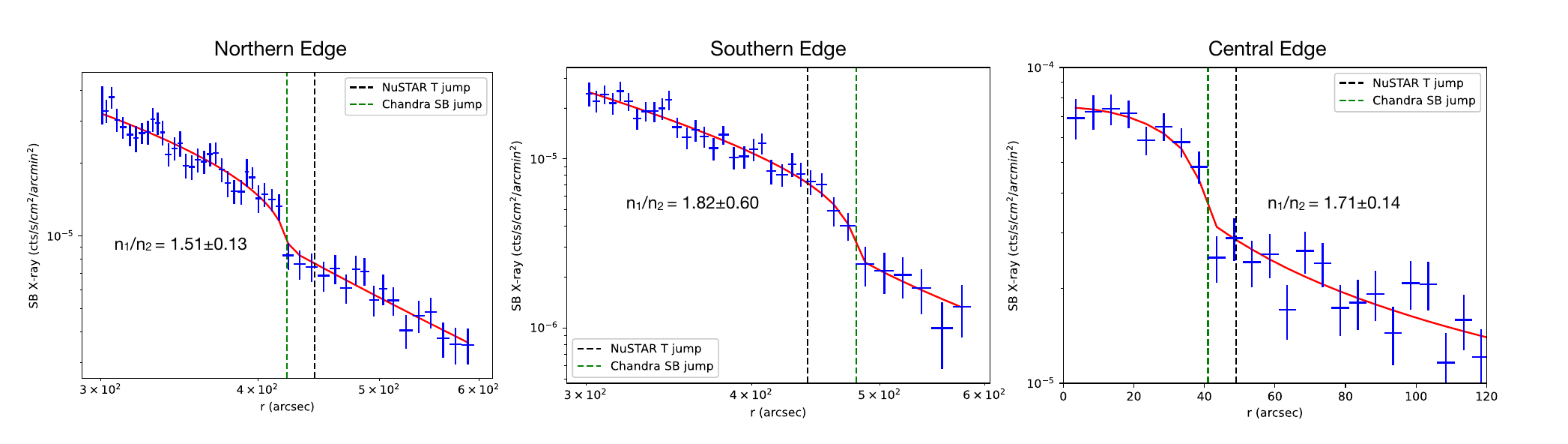}
\caption{\chandra~X-ray surface brightness profiles (blue) and the best-fit model (red) of northern, southern and central SB edges. \label{fig:SBjumpChandra}}
\end{figure*}

We then studied the \chandra~SB profiles of the SB edges evidenced by \nustar, \chandra, LOFAR images, as well as our GGM analysis and temperature maps. These regions are 3~-~4 (C$_{in}$~-~C$_{out}$), 7~-~8 (N$_{in}$~-~N$_{out}$), and 9~-~10 (S$_{in}$~-~S$_{out}$), where the first region in these region couples are presumed post-shock and the second one represents the pre-shock.

Although \citet{zhang20} study the SB profiles of northern and southern SB edges, we extracted new \chandra~ SB profiles from the exact regions used in this work for precision. The profiles and the best fit model are shown in Fig.~\ref{fig:SBjumpChandra}. Assuming the adiabatic index of an ideal monoatomic gas of $\gamma$=5/3, Mach numbers obtained from the density jumps are, $\mathcal{M}$~=~1.35, 1.58, and 1.50 for the northern, southern and central edges, respectively. The Mach numbers calculated by the temperature jumps from our crosstalk analysis range between 1.22~-~1.37. In addition, we calculated the Mach number for the temperature jump between regions 6-8, accounting for the possibility that ``hot spot" is truly a hot plasma where the post-shock is localized at Region 6, i.e., a bullet-like feature, and Region 8 is the pre-shock region. This approach results in a $\mathcal{M}$~$\sim$~2. These Mach numbers are presented in Table~\ref{tab:mach}.

The details of the {\tt nucrossarf} analysis are presented in Section~\ref{sec:crossarfapp}.

\subsection{Local IC search} \label{sec:IC} 

\begin{figure}[h!]
\centering
\includegraphics[width=60mm]{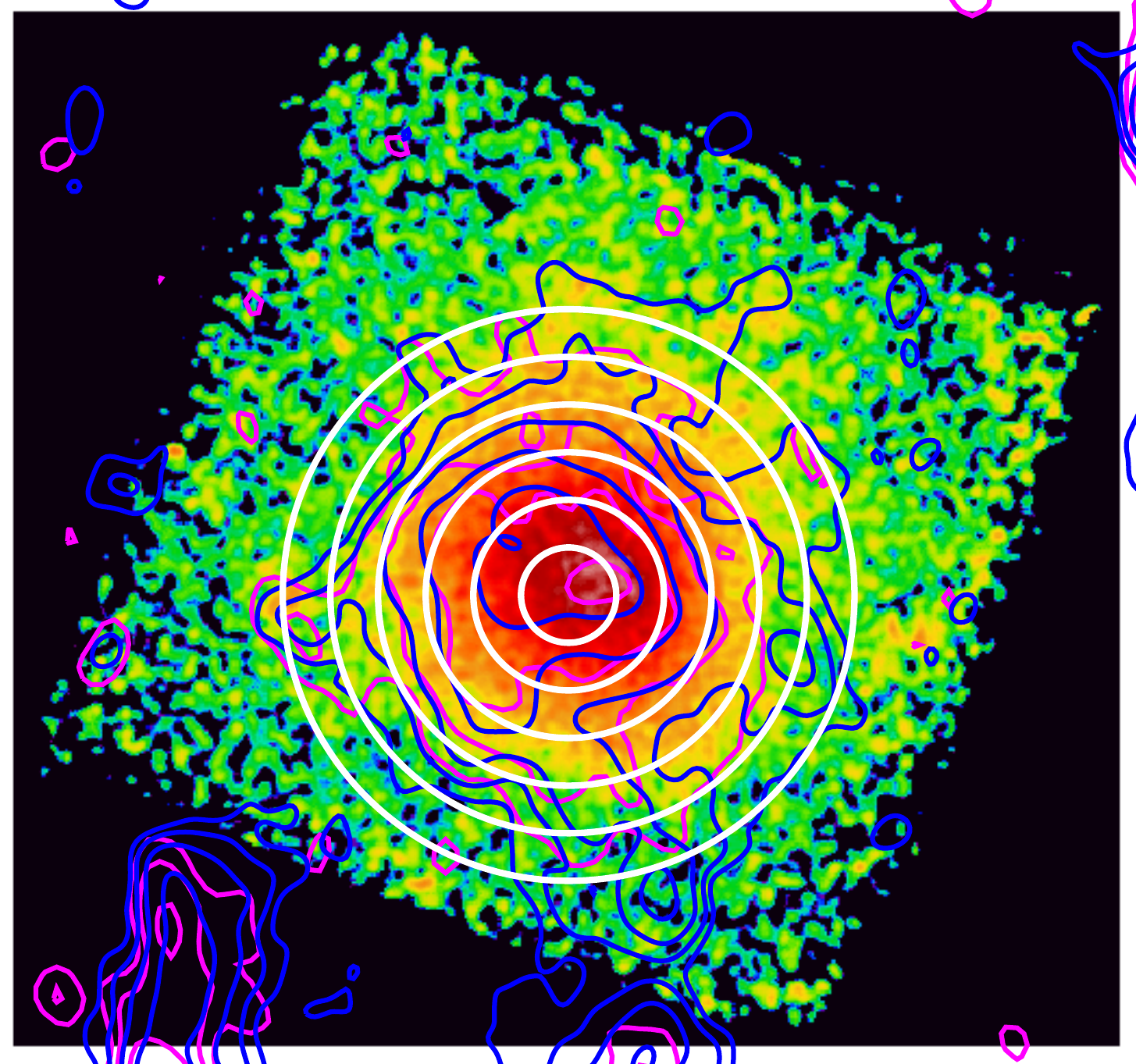}
\caption{Selected regions (white annuli) for IC search overlaid on \nustar\ 3.0~-~10.0 keV image. VLA L-band D configuration (magenta, at 3$\sigma_{rms}\times$ [1, 2, 4, 8] levels, where ($\sigma_{rms}$ = 70~$\mu$Jy/beam), and LOFAR (blue, at $\sigma_{rms}\times$ [3, 6, 12, 24, 48]. ($\sigma_{rms}$ = 330~$\mu$Jy/beam [46$\arcsec\times$45$\arcsec$]) radio contours are also shown.\label{fig:IC}}
\end{figure}

We did not find strong evidence for extended IC emission from our global spectra fit results, but we continued to search for IC locally within the radio halo, with the motivation that IC may start to dominate the emission in narrower regions moving away from the center of the cluster where ICM dominates. For this analysis, we used annular regions, where \citet{hoang21} studies radio spectral indices and we extracted \nustar~spectra from 5 annuli as shown in Fig.~\ref{fig:IC}. Although 6 regions are selected by \citet[Fig.~7]{hoang21} including a central circular region, we exclude that central region from our analysis expecting for the ICM to dominate the total emission, hence we studied 5 annular regions. We applied the same analysis done for the \nustar~global spectra to these annular spectra. We also allowed for metal abundances to be free. Model parameter results are shown in Table~\ref{tab:ICfit}.

Annulus 1 spectrum is best described by a single temperature model, whereas for annuli 2~-~3~-~4, 2T model describes the spectra the best-fit models. The outermost annulus, annulus 5, is best described by 1T+IC model, but almost equally well as the 2T model does. This region roughly passes through the Regions 8 and 10 of our crosstalk analysis.

\begin{deluxetable*}{ccccccccc}
\tabletypesize{\scriptsize}
\tablewidth{0pt} 
\tablecaption{IC search sprectral fit results from selected annuli shown in Fig.~\ref{fig:IC}. Model norms are described in Table~\ref{tab:globalfit}. IC flux is given for the 20.0~-~80.0 keV band in 10$^{-13}$~erg~s$^{-1}$~cm$^{-2}$.
\label{tab:ICfit}}
\tablehead{
\colhead{Annulus}& \colhead{} &\colhead{$kt$} &\colhead{\textit{Z$_{1}$}}& \colhead{{\it norm}} &\colhead{$kt$ or $\Gamma$} & \colhead{{\it norm} or $\kappa$} & \colhead{IC} &\colhead{}\\[-0.95em]
\colhead{Number}& \colhead{Model} &\colhead{(keV)} &\colhead{({\it Z$_{\odot}$})}& \colhead{(10$^{-3}$)} &\colhead{keV or \nodata} & \colhead{(10$^{-3}$~cm$^{-5}$) or (10$^{-4}$)} & \colhead{Flux} &\colhead{{\it C / $\nu$}}
}
\startdata
\\[-0.95em]
1 & 1T & 10.28$^{+0.21}_{-0.20}$& 0.31$\pm{0.06}$& 2.808$^{+0.062}_{-0.061}$   & \nodata & \nodata & \nodata & 1218.21 / 1318 \\
\\[-0.95em]
 & 2T & 10.34$^{+0.23}_{-0.21}$& 0.33$^{+0.07}_{-0.06}$& 2.786$^{+0.065}_{-0.067}$   & 0.16$^{+0.27}_{-0.13}$ & 5.390E7$^{+1.572E10}_{-5.401E7}$ & \nodata  & 1216.92 / 1316 \\
 \\[-0.95em]
 & 1T + IC & 10.28$^{+0.25}_{-0.28}$ & 0.31$\pm{0.07}$& 2.373$^{+0.102}_{-0.231}$   & 2.10 (fixed) & 0.300$^{+0.880}_{-0.300}$ & 0.46 & 1218.21 / 1317 \\
  \\[-0.95em]
 \hline
\\[-0.95em]
2 & 1T & 9.42$^{+0.21}_{-0.20}$& 0.31$\pm{0.06}$& 2.846$^{+0.069}_{-0.068}$   & \nodata & \nodata & \nodata & 1385.03 / 1365 \\
\\[-0.95em]
 & 2T & 10.34$^{+0.91}_{-0.54}$& 0.36$\pm{0.07}$& 2.527$^{+0.178}_{-0.371}$   & 2.33$^{+2.13}_{-1.02}$ & 0.915$^{+1.235}_{-0.423}$ & \nodata  & 1380.11 / 1363 \\
 \\[-0.95em]
 & 1T + IC & 8.90$^{+0.37}_{-0.43}$ & 0.38$^{+0.08}_{-0.07}$& 2.384$^{+0.239}_{-0.240}$   & 2.10 (fixed) & 1.853$^{+0.937}_{-0.934}$ & 2.85 & 1381.48 / 1364 \\
  \\[-0.95em]
 \hline
\\[-0.95em]
3 & 1T & 8.80$^{+0.26}_{-0.27}$& 0.13$\pm{0.06}$& 2.165$^{+0.066}_{-0.065}$   & \nodata & \nodata & \nodata & 1321.22 / 1351 \\
\\[-0.95em]
 & 2T & 12.91$^{+2.71}_{-2.78}$& 0.22$^{+0.11}_{-0.08}$& 1.371$^{+0.329}_{-0.294}$   & 2.87$^{+1.30}_{-1.64}$ & 1.689$^{+1.809}_{-0.334}$ & \nodata  & 1303.74 / 1349 \\
 \\[-0.95em]
 & 1T + IC & 6.84$^{+0.68}_{-0.60}$ & 0.24$^{+0.09}_{-0.08}$& 1.464$^{+0.163}_{-0.162}$   & 2.05 (fixed) & 2.842$^{+0.611}_{-0.610}$ & 5.25 & 1306.49 / 1350 \\
  \\[-0.95em]
 \hline
\\[-0.95em]
4 & 1T & 7.34$^{+0.28}_{-0.27}$& 0.05$^{+0.06}_{-0.05}$& 1.897$^{+0.079}_{-0.078}$   & \nodata & \nodata & \nodata & 1329.64 / 1361 \\
\\[-0.95em]
 & 2T & 7.70$^{+0.39}_{-0.33}$& 0.09$^{+0.07}_{-0.06}$& 1.778$^{+0.094}_{-0.102}$   & 0.39$^{+0.25}_{-0.16}$ & 1.271E4$^{+2.378E8}_{-1.319E4}$ & \nodata  & 1324.24 / 1359 \\
 \\[-0.95em]
 & 1T + IC & 7.23$^{+0.37}_{-0.59}$  & 0.05$^{+0.06}_{-0.05}$& 1.837$^{+0.89}_{-0.215}$   & 2.10 (fixed) & 0.231$^{+0.841}_{-0.231}$ & 0.36 & 1329.53 / 1360 \\
  \\[-0.95em]
 \hline
\\[-0.95em]
5 & 1T & 7.38$\pm{0.39}$ & 0.12$\pm{0.08}$& 1.491$^{+0.087}_{-0.079}$   & \nodata & \nodata & \nodata & 1308.92 / 1435 \\
\\[-0.95em]
 & 2T & 24.40$^{-25.08}_{-15.63}$& 0.17$^{+0.12}_{-0.08}$& 0.348$^{+0.722}_{-0.193}$   & 4.62$^{+1.35}_{-2.57}$ & 1.457$^{+0.208}_{-0.546}$ & \nodata  & 1304.14 / 1433 \\
 \\[-0.95em]
 & 1T + IC & 5.55$^{+1.01}_{-1.14}$ & 0.26$^{+0.19}_{-0.13}$& 0.975$^{+0.228}_{-0.198}$   & 2.15 (fixed) & 2.300$^{+0.798}_{-0.943}$ & 2.94 & 1304.25 / 1434 \\
  \\[-0.95em]
\enddata
\end{deluxetable*}

\section{Summary and Discussion}\label{sec:discussion}

\cl~ hosts a giant radio halo and double radio relics pointing to a late stage merger. The location of radio relics points to a merger plane perpendicular to the line of sight. Motivated by the SB edges detected by \chandra~\citep{zhang20} at the central region of the cluster, we observed the cluster with \nustar~ with the aim of studying the nature of these SB edges. We applied spectro-imaging methods using \nustar~and \chandra~data, and relying on the recent LOFAR study by \citet{hoang21}. 

\subsection{Global view}
We first studied the photon images of \nustar, \chandra~and LOFAR data. To capture any strong SB gradients, we used GGM filter on the \nustar~ photon images and we saw that an interesting SB edge feature consistently appears at the central region of the cluster with \nustar, \chandra~and LOFAR images as well as \nustar~GGM filter.

Background extracted, exposure corrected images (Fig.~\ref{fig:nuphoton}) of \nustar~and \chandra~data point to a localized, enhanced emission at the central region, where we assigned Region 3 to, for our crosstalk analysis (Fig.~\ref{fig:crossarfmask}). Furthermore, GGM and LOFAR images suggest a sharp edge between Regions 3 and 4, and the correspondence of this X-ray edge with the radio band, respectively, as seen in Fig.~\ref{fig:inneredge}. 

In addition, \nustar~image at hard (10.0~-~20.0 keV) energy band hints at high temperature or the existence of non-thermal emission enclosed by the SB edges suggested by \citet{zhang20} using \chandra~ data. We found that the this hard excess is due to the hot ICM, given that we did not find strong IC component. We do not directly detect a clear SB depression at the proposed \citep{zhang20} ``Channel" region from \nustar~images.

The offset between the diffuse radio and X-ray central halo peak seen in the photon images (Fig.~\ref{fig:nuphoton}), as well as the \nustar~SB and radio contours overlaid on the temperature map (Fig.~\ref{fig:TempMap}), is also observed by \citet{brown11b} for this cluster, as well as in Coma cluster \citet{deiss97}. This is thought to be due the non-equilibrium state of the merger system \citep{brown11b}.

The global analysis of the cluster constitutes the selection of a circular region (5.2$\arcmin$) containing the radio halo (roughly enclosing Regions 1-9), extraction of spectra of this region from both \nustar~and \chandra~data, and fitting the spectra with single temperature, two temperature and single temperature plus non-thermal emission models. The purpose of this analysis is to assess the global properties of the cluster and to search for possible extended IC emission in the cluster. Assuming an isothermal plasma, the global temperature found by \nustar~is {\it kT} = 9.1$\pm{0.1}$ keV, and for joint \nustar~and \chandra~fit, we find {\it kT} = 9.2$\pm{0.3}$ keV. The results of \nustar, and joint \nustar~and \chandra~data analyses show that the statistics improve with an additional {\tt apec} or {\tt powerlaw} component, with respect to a single temperature {\tt apec} model. The {\tt powerlaw} component was more dominant in \nustar~fit than joint \nustar~and \chandra~fit, due to the band pass of \chandra~being narrower than that of \nustar. Therefore, joint fit captures more of the thermal emission. Two temperature model for the \nustar~ fit has a high temperature component of T$_{H}$ = 10.6$^{+0.8}_{-0.4}$ keV and a lower temperature component with T$_{L}$ = 2.2$^{+1.1}_{-0.5}$ keV. However, our further detailed analyses on the temperature structure, i.e., temperature map and crosstalk analysis, do not show signs of dominant emission around this low temperature value. Apparently, the multi-temperature structure of the cluster would theoretically be better described with a multi-temperature model, yet the temperature values of various regions within the radio halo being high and close to each other makes it statistically difficult to disentangle these components. 

In the joint \nustar~and \chandra~fit, however, we witnessed a subtle high temperature component of T$_{H}$ = 21.56 $^{+2.71}_{-2.90}$ keV and a dominant low component of T$_{L}$ = 6.4$\pm{0.4}$ keV. The lower limit found for Region 6 for the hot spot scenario in the crosstalk analysis results, is actually within 1$\sigma$ of this global high temperature component. The lower component, however, is too low with respect to what we see in the central region throughout this work, and also with respect to the findings of \citet[Table 1]{zhang20}. Overall, the statistics for both \nustar, and joint \nustar~and \chandra~fit improved by adding a second {\tt apec} model with {\it $\Delta$C/$\Delta\nu$} = 41.25/2 and {\it $\Delta$C/$\Delta\nu$} = 36.96/2, respectively. 

The addition of a {\tt powerlaw} component to the single temperature model improved the statistics by {\it $\Delta$C/$\Delta\nu$} = 38.3/1 and {\it $\Delta$C/$\Delta\nu$} = 32.3/1 for \nustar, and joint \nustar~and \chandra~analysis, respectively. Statistically a 2T fits the data better than a 1T+IC description (Table~\ref{tab:globalfit} and Table~\ref{tab:globalfitNuCh}). The temperature component of the 1T+IC model is $\sim$8 keV. \nustar, and joint \nustar~and \chandra~analysis results agree within 1$\sigma$. 

Although the data are better described by the 2T model, we provide an IC flux only to report an upper limit, as if IC were detected. The {\tt powerlaw} component was more dominant in the \nustar~fit than that of the  joint fit. Therefore, we use \nustar~fit results for our flux calculations. The upper limit to the IC flux in the 20.0~-~80.0 keV band was calculated using the {\tt cflux} model applied to the {\tt powerlaw} of the \nustar~fit. The resulting flux is 2.695~$\times$~10$^{-12}$~erg~s$^{-1}$~cm$^{-2}$. Using the LOFAR data \citep{hoang21}, this refers to a lower limit of ~0.08$\mu$G for the average magnetic field.

We also studied the effect of N$_{H}$ parameter on our global spectra results using priors from \citet{zhang20}. Since our global region radius is $\sim$300\arcsec, the only prior we could use for N$_{H}$ was fixing its value to 8.21$\times10^{21}$~cm$^{-2}$ suggested by \citet{zhang20}, to test its impact on the global temperature and IC limit. By fixing the N$_{H}$ to this value during the joint \nustar~and \chandra~1T model fit, statistics become 2120.48/2108 d.o.f., where with free N$_{H}$, it is 2104.29/2107 d.o.f. Between these two approaches, the global temperature varies by 0.8\%. And for the joint \nustar~and \chandra~1T+IC model fit, fixing the N$_{H}$ decreases the magnetic field lower limit by only 5\% as opposed to letting it vary.

In addition, we studied the effect of the photon index, $\Gamma$, on the IC flux using our global joint \nustar~and \chandra~analysis (1T+IC model). We varied $\Gamma$ between 1.9 and 2.1 from the fixed value of $\Gamma$ = 2.0. $\Gamma$ = 2.0 is obtained from the spectral index ($\alpha$) value in \citet{hoang21}. The uncertainty given by \citet{hoang21} on this $\alpha$ is +/-0.05, and here we use a more conservative approach by taking the error as +/-0.10. The IC flux changes by 9\% and 12\% for setting $\Gamma$ = 1.9 and $\Gamma$ = 2.1, respectively, instead of $\Gamma$ = 2.0.

\subsection{The ``hot spot"}\label{sec:hotspotdiscussion}
Our temperature map in Fig.~\ref{fig:TempMap} shows a hot spot ($\sim$12 keV) at the location of the northern SB edge. We extracted \nustar~spectra from this region to study the nature of the emission. We propose three scenarios to explain the emission from this source, which are (1) a hot spot where the shock heated gas is concentrated at that region (bullet-like), (2) a highly obscured AGN, and (3) a localized IC emission connected to the radio source in the vicinity. 

Shocks driven by a piston-like object, like a cool core, do not produce shocks with uniform Mach number across it, but will have higher Mach numbers near the piston and lower farther away. If the ``hot spot" is actually of thermal origin, this would be a similar case as in the bullet cluster. If the shock surface is considered as a 2D sheet, the middle will have the highest Mach number (in this case $\mathcal{M}$~=~2), and it will be lower everywhere else. At the hot spot-Region 8 interface the temperature jump will be highest and the particle acceleration will be most efficient. 

Supported by theoretical and high-resolution hydrodynamical simulations (\citealt{gaspari17,gaspari19}), AGN in massive hot halos often become heavily obscured via the CCA triggered by the top-down multiphase condensation of warm and cold clouds, which rain onto the central supermassive black hole. Such obscuring CCA rain has been also probed in lower energy bands, e.g., with ALMA and MUSE absorption/emission-line features (\citealt{rose19,olivares22,temi22}). 

However, at the ``hot spot" location, there are no known AGNs. We also carried out a simple mid-IR selection criteria following \citep{taweewat22} using AllWISE2020 Catalog \footnote{\url{https://irsa.ipac.caltech.edu/Missions/wise.html}}. Within r=1$\arcmin$ of the ``hot spot" position, we found an IR object with ID J021703.24+703909.0
and found W1-W2=-0.598. This value is much lower than what is expected from an AGN emission \citep{taweewat22}. However we note that, mid-IR selection for AGN mostly select extremely bright AGN (high accretion rate). This implies that the object may not be a strong AGN, but could still be a fainter/smaller AGN.

Furthermore, the vicinity of the radio galaxy may indeed be causing the IC emission due to the reacceleration of relativistic particles injected from the site, or by adiabatic compression, the merger shock may be ``re-energizing" the radio plasma. The radio galaxy spectral index found by LOFAR is within 1$\sigma$ of the IC emission photon index we found.

Our \nustar~analysis using {\tt nuproducts} shows that all three scenarios are almost equally likely as represented in Table~\ref{tab:hotspotscenario}. However, 2T model fit to the joint \nustar~and \chandra~data revealing a similar high temperature (Table~\ref{tab:globalfitNuCh}) for the global assessment with the crosstalk results for Region 6 of the hot spot (Table~\ref{tab:nuprocrossfit}) may not be a coincidence. Although we would expect \nustar~to capture the higher temperature component model better with respect to \chandra, the addition of the \chandra~data may have helped better constraining the lower temperature model. Since \chandra~data are quite shallow and foreground absorption is moderately high, it is difficult to make a conclusion. 

\subsection{Crosstalk analysis and Shock fronts}
To account for the scattering due to \nustar~PSF, we applied {\tt nucrossarf} to dissociate the emission originating from a specific region, from the other regions in the field of view. Our crosstalk analysis (Table~\ref{tab:nuprocrossfit}) reveal several temperature jumps between the regions selected with the guidance of \nustar, \chandra~and LOFAR images, as well as temperature and GGM maps. These weak jumps occur at Region 7-8 (northern) and 9-10 (southern), and Region 3-4 (central) interface. We calculated Mach numbers for these jumps, and found $\mathcal{M}$~=~1.22~-~1.37 (Table~\ref{tab:mach}), hinting at multiple weak shocks within the radio halo. Our \chandra~SB profiles show that these temperature jumps have corresponding density jumps with $\mathcal{M}$~=~1.35~-~1.58. The Mach numbers derived from the \nustar~temperature jumps and \chandra~density jumps found in this work are consistent with each other. The Mach number obtained for the density jumps from the southern SB edge detected by \citet{zhang20} is 2$\sigma$ higher than what we find in this work. We reason that the shock may be the strongest (localized) at the center when a smaller angle sector is selected for the analysis, as we similarly discuss in the case of the hot spot scenario in Section~\ref{sec:hotspotdiscussion}.

The physical distances of the temperature (density) jumps are $\sim$500 ($\sim$440) kpc, $\sim$650 ($\sim$780) kpc, and $\sim$100 ($\sim$72) kpc from the emission peak center for the northern, southern and central fronts, respectively. The difference between the locations of the temperature-density jumps may be due to the imperfect astrometry of \nustar, as well as the poor photon statistics of the \chandra~data. Deeper \chandra~observations could reveal more accurate and precise locations of the density jumps aided by \chandra~GGM analysis followed by various deprojected radial SB profiles across the FOV.

On the technical side, we present a new code in this work, {\tt nucrossarf}, aimed at accounting for the X-ray scattering in and out of spectral extraction regions within the \nustar~FOV. We ran the code for three different scenarios for 17 sources, where the temperature and model normalization values of all sources are almost identical for the scenarios that points to the stability of the code. Furthermore, {\tt nucrossarf} and {\tt nuproducts} temperature values are in agreement within 1$\sigma$ for Regions 1-2-4-5-8-9 and 2$\sigma$ for Regions 3-7-10. Temperature values for Regions 11-12-13, that are at the outskirts of the cluster drop by $\sim$2 keV, pointing to a contamination of hard excess from the points sources enclosed by them, which is supported by the fact that the normalizations of these point source emissions are reduced by a factor of 2 after the crosstalk treatment. Regions 6- the ``hot spot", and Region 14 -the Channel temperatures from {\tt nuproducts} are quite different from {\tt nucrossarf} results. 

For the ``Channel" region, we see that the temperature and normalization parameters were poorly constrained. The scattered emission from the neighbouring regions may be responsible for this instability, hinting at a very faint emission at the Channel region.

The total fit C statistics for the crosstalk analyses are {\it C/$\nu$} = 14669.14/15218, {\it C/$\nu$}= 14694.08/15219, {\it C/$\nu$} = 14666.06/15218 for hot spot, obscured AGN, and IC scenarios that were assumed to explain the Region 6 emission, respectively. Although the best fit is obtained for the IC scenario, at this depth, all scenarios are almost equally likely.

\subsection{IC search}

We also carried out a local search for IC for regions where radio spectral indices are given by \citet{hoang21} inside the radio halo. The outermost annulus fit results show that 1T+IC model describes the spectra the best, but almost equally well with 2T model. For this region, we find an upper limit of 2.64~$\times$~10$^{-13}$~erg~s$^{-1}$~cm$^{-2}$ in the 20.0~-~80.0 keV band for the IC flux, if were detected. This region where 1T+IC model and 2T models have quite similar statistics, passes through Region 8 and 10 that is the outer region to the northern and southern shock fronts. Whereas annulus 3 has the highest IC flux, again has similar statistics with 2T model, and coincides with the northern and southern SB edges, that is expected due to shock enhancement \citep[see, e.g.]{sarazin02}.

\section{Conclusion}\label{sec:conclusion}

Based on the \nustar~and \chandra~data, complemented by a previous LOFAR study \citep{hoang19}, we studied the late stage galaxy cluster merger \cl~. We present our conclusions of this work in this section.

All three scenarios, namely; hot spot, obscured AGN and localized IC emission, proposed for the ``hot spot" captured by our temperature map seem to be equally likely. If the ``hot spot" is indeed a thermal plasma of $\sim$12 keV, it indicates a shock front $\sim$500 kpc away from the cluster emission peak. For the obscured AGN or IC emission scenarios deep optical and deep low frequency radio observations are required to study the galaxy distribution at the site. In essence, what is witnessed here may be even a combination of a hot spot + IC.

We confirm two shock fronts at Region 7-8 (northern) and 9-10 (southern) suggested by \citet{zhang20}, and detect another shock front at Region 3-4 interface (central), within 0.5 r$_{500}$ of the cluster. The physical distances of the temperature (density) jumps are $\sim$500 ($\sim$440) kpc, $\sim$650 ($\sim$780) kpc, and $\sim$100 ($\sim$72) kpc from the emission peak center for the northern, southern and central fronts, respectively. These weak shocks may be partially responsible for sustaining the giant radio halo emission by the (re)acceleration of the relativistic particles at the central region of the cluster. 

The axis connecting these northern and southern secondary shock fronts are almost perpendicular to the axis connecting the radio relics, suggesting that these secondary shocks are unrelated to the first core passage. The coincidence of the northern and southern shock fronts with the radio halo emission edge adds this cluster a short list of clusters that suggests the radio halos formation may be boosted by the turbulence formed directly behind shocks. Deeper high angular resolution X-ray observations should reveal $\sim$ 100 kpc scale turbulence eddies, taking advantage of the merger plane lying perpendicular to the line of sight.

To the best of our knowledge, this cluster is the second known case where a secondary shock coinciding with the radio halo emission in the form of subsequent settling of the ICM is observed within 1 Mpc, the first being Coma \citep{simionescu13}. 

Unlike Coma, \cl~shows multiple shock fronts, which makes this cluster the {\it only} known case where multiple secondary shock fronts are formed within 0.5~r$_{500}$. We may have captured this cluster before the settling of the ICM, hence, further studies beyond local systems are needed to be the carried out to investigate the ubiquity of these secondary shocks to understand the evolution of galaxy cluster mergers and radio halo formation mechanisms. Galaxy clusters hosting giant radio halos can be the first place to look for similar cases.

\acknowledgments
We thank the anonymous referee for valuable suggestions that improved our manuscript. AT \& DRW acknowledges support from NASA NuSTAR GO grant 80NSSC22K0066 and from NASA ADAP award 80NSSC19K1443. MG acknowledges partial support by NASA Chandra GO9-20114X and {\it HST} GO-15890.020/023-A, and the {\it BlackHoleWeather} program. RJvW acknowledges support from the ERC Starting Grant ClusterWeb 804208. CS acknowledges support from the MIUR grant FARE ``SMS". AT thanks Ross Silver and Taweewat Somboonpanyakul for valuable discussions. This research has made use of data from the \nustar\ mission, a project led by the California Institute of Technology, managed by the Jet Propulsion Laboratory (JPL), and funded by by the National Aeronautics and Space Administration (NASA). In this work, we used the NuSTAR Data Analysis Software (NuSTARDAS) jointly developed by the ASI Science Data Center (ASDC, Italy) and the California Institute of Technology (USA). The data for this research have been obtained from the High Energy Astrophysics Science Archive Research Center (HEASARC), provided by NASA’s Goddard Space Flight Center. This research has also made use of data obtained from the Chandra Data Archive and the Chandra Source Catalog, and software provided by the Chandra X-ray Center (CXC) in the application package CIAO.

\bibliography{CL0217}{}
\bibliographystyle{aasjournal}

\appendix

\section{\nustar\ Background assessment and systematics}\label{sec:bgdfit}

\begin{figure*}[h!]
\centering
\includegraphics[width=170mm]{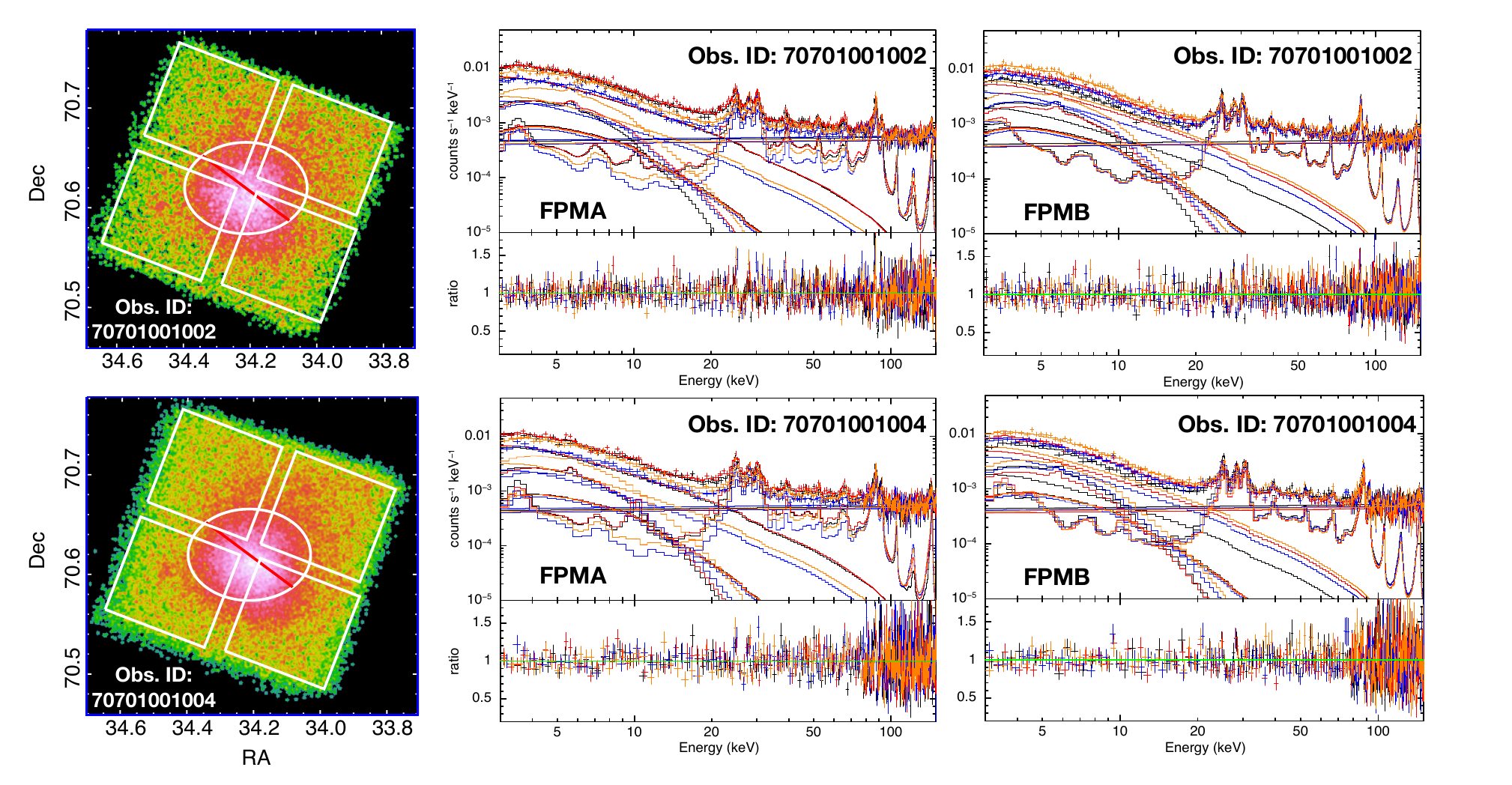}
\caption{Selected regions for background analysis (\textit{left panel}) laid over cleaned \nustar~3-10 keV photon images. White squares indicate regions where the background spectra is extracted. Joint-fit of background and cluster emission of \nustar\ FPMA (\textit{middle panel}) and FPMB (\textit{right panel}) for Obs. ID 70701001002 (\textit{upper panel}) and Obs. ID 70701001004 (\textit{lower panel}). Each color represents a region selected for the background fit. For plotting purposes, adjacent bins are grouped until they have a significant detection at least as large as 15$\sigma$, with maximum 15 bins. \label{fig:nuskybgd}}
\end{figure*}

In this section, we present the spectra from the background model fits using both FPMs and both Obs. ID in Fig.~\ref{fig:nuskybgd}. As shown in the left panel of Fig.~\ref{fig:nuskybgd}, we selected 4 regions in the FOV from each detector chip, excluding the ICM emission as much as possible but also including as much data as available. Region selection is somewhat an experienced guess, and we tested the fits for smaller and larger regions to optimize the stability of the fit. The method is described in \citet{wik14}.

In addition, we note that single temperature {\tt apec} model is sufficient to account for local and scattered cluster emission inside background regions, and more complicated models (e.g., two temperature {\tt apec} model) do provide significant improvement. More importantly, the additional component, if not limited by strict priors, will generally try to model some other feature of the ICM spectra where the background model is not perfect.

To quantify the systematics of the fCXB in the \nustar~ background analysis \citep{wik14}, we chose Region 5 that has the least complicated thermal structure. We increased and decreased the fCXB component of the background by 50\% in the spectral fits to see its effect on the temperature measurements. We chose 50\% since, given the cosmic variance for a region of this size, it is the expected variation in CXB that is also comparable to the brightness of the detected point sources. Increasing the fCXB component of the background by 50\% resulted in a temperature decrease of 0.04 keV, whereas decreasing the fCXB component of the background by 50\% resulted in a 0.06 keV increase in the temperature. These values correspond to less than 35\% of the statistical errors on the temperature of that region.

The dominant systematic uncertainty of the \nustar~background comes from the aperture cosmic X-ray background (aCXB) characterization, which dominates the $E\sim$10~keV background, but it is generally too flat to bias the thermal continuum and thus temperature measurements \citep{wik14}. This statement is less true when an IC model is included, but since the data do not support IC emission over pure thermal emission, the effect of this systematic uncertainty is negligible concerning our results.

\begin{figure*}[h!]
\includegraphics[width=180mm]{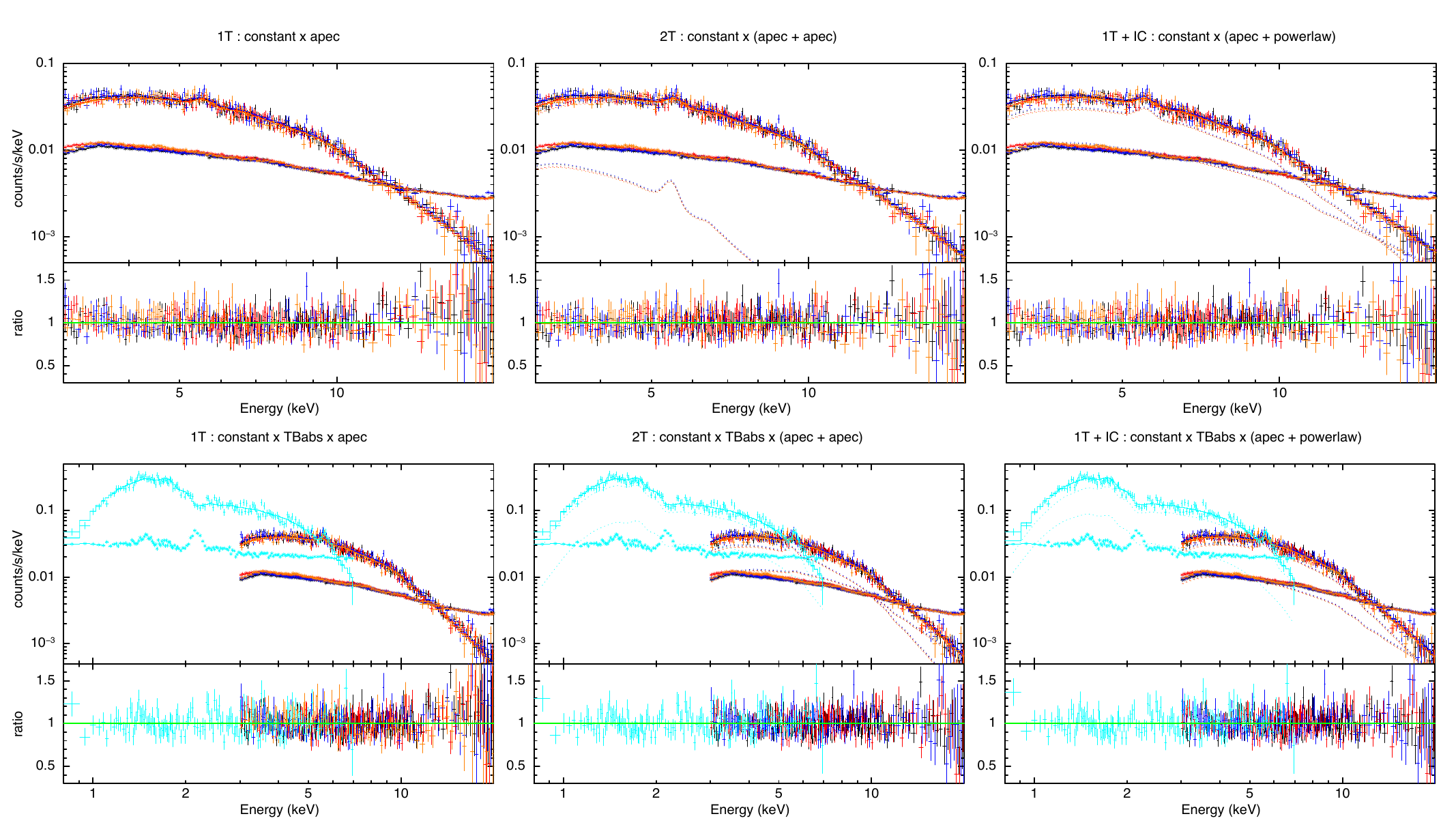}
\caption{Global fits of \nustar\ (\textit{upper panel}), and joint \nustar\ and \chandra\ data (\textit{lower panel}) with 1T (\textit{left panel}), 2T (\textit{middle panel}), and 1T + IC (\textit{right panel}) of the spectra extracted from the central 5$\arcmin$.2 circular region. Black indicates FPMA and red indicates FPMB for ObsID 70701001002, whereas for ObsID 70701001004, FPMA and FPMB are indicated by blue and orange, respectively. \chandra\ ACIS-I is indicated by turquoise. The data and models are shown as the higher curve and lower lines correspond to background spectrum. The dashed curves correspond to the model components to visualize their contribution to the composed model. The ``ratio” panel shows data to model ratios describing the goodness of the fit. For plotting purposes, adjacent bins are grouped until they have a significant detection at least as large as 8$\sigma$, with maximum 12 bins. \label{fig:globalNuCh}}
\end{figure*}

\section{\nustar\ and \chandra\ global fit spectra}\label{sec:globalfit}

We present the results of the global spectral fit in this section for only \nustar~and joint \nustar~and \chandra~data using 1T, 2T, and 1T + IC models as shown in Fig.~\ref{fig:globalNuCh}. Since the statistics are quite similar, the differences in ratios are not discernible, yet the model contribution curves provide an insight to the contribution of different models.

\section{``hot spot" scenarios}\label{sec:hotspot}

\begin{figure*}[h!]
\includegraphics[width=162mm]{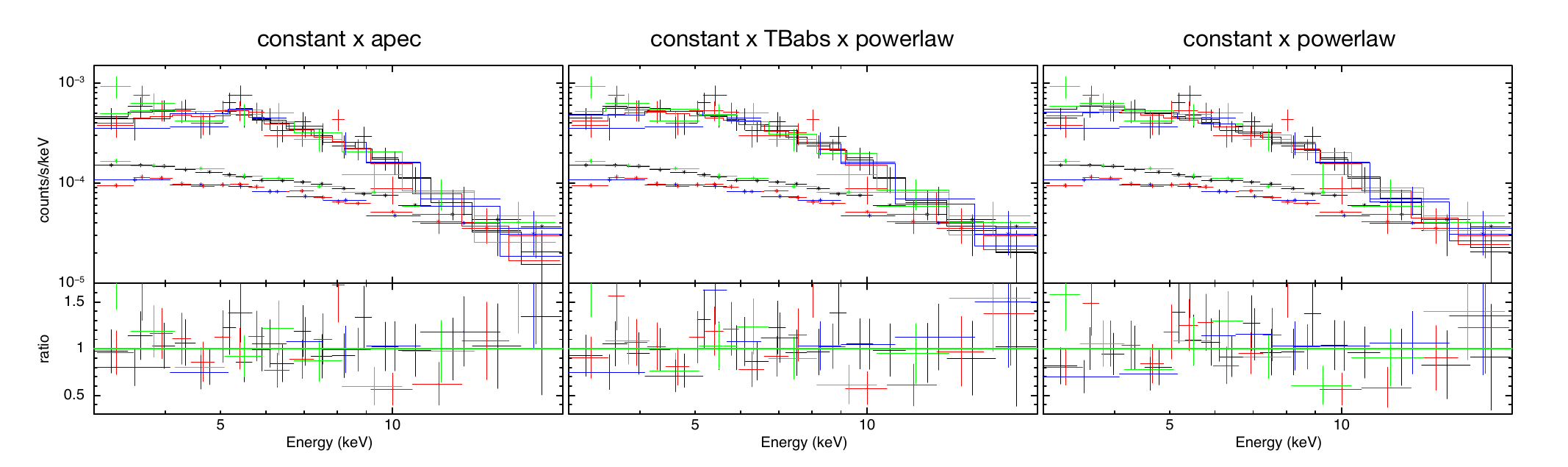}
\centering
\caption{Spectra of the \nustar\ ``hot spot" region fits with {\tt constant} $\times$ {\tt apec} (\textit{left panel}), {\tt constant} $\times$ {\tt TBabs} $\times$ {\tt powerlaw} (\textit{middle panel}), and {\tt constant} $\times$ {\tt powerlaw} (\textit{right panel}).  Black indicates FPMA and red indicates FPMB for ObsID 70701001002, whereas for ObsID 70701001004, FPMA and FPMB are indicated by green and blue, respectively. For plotting purposes, adjacent bins are grouped until they have a significant detection at least as large as 4$\sigma$, with maximum 8 bins. \label{fig:hotspotscenario}}
\end{figure*}

The resulting spectra from the three scenarios proposed for the explanation of the Region 6 emission is presented in this section. The statistics are again similar, which makes it difficult to conclude from the figures which model best describes the data. However, in its spectrum we encountered a strange dip around 4.8 keV in Fig.~\ref{fig:hotspotscenario}. We extracted a spectra from a larger (1$\arcmin$) circular region at the location at the ``hot spot", as well as different data and plotting bins. The dip feature still persists for those spectra as well, pointing to a reason different from a mere coincidence of statistical fluctuations. We could not found an explanation for this feature, and we refrain from any speculations.

\section{nucrossarf}\label{sec:crossarfapp}

With this detailed study, we also happen to test the new {\tt nucrossarf} code. In this section, we elaborate the inner workings of the code, and what to expect from the code for future users.

{\tt nucrossarf} mainly calculates the cross-ARFs of a number of user defined extracted regions, and dissociates the source distribution from the contamination from other regions by jointly fitting the spectra of all regions. For N extracted regions, {\tt nucrossarf} generates {\tt N} $\times$ {\tt N} ARFs to account for the wings of the PSF of each source present in other regions. 

\begin{figure*}[h!]
\centering
\includegraphics[width=140mm]{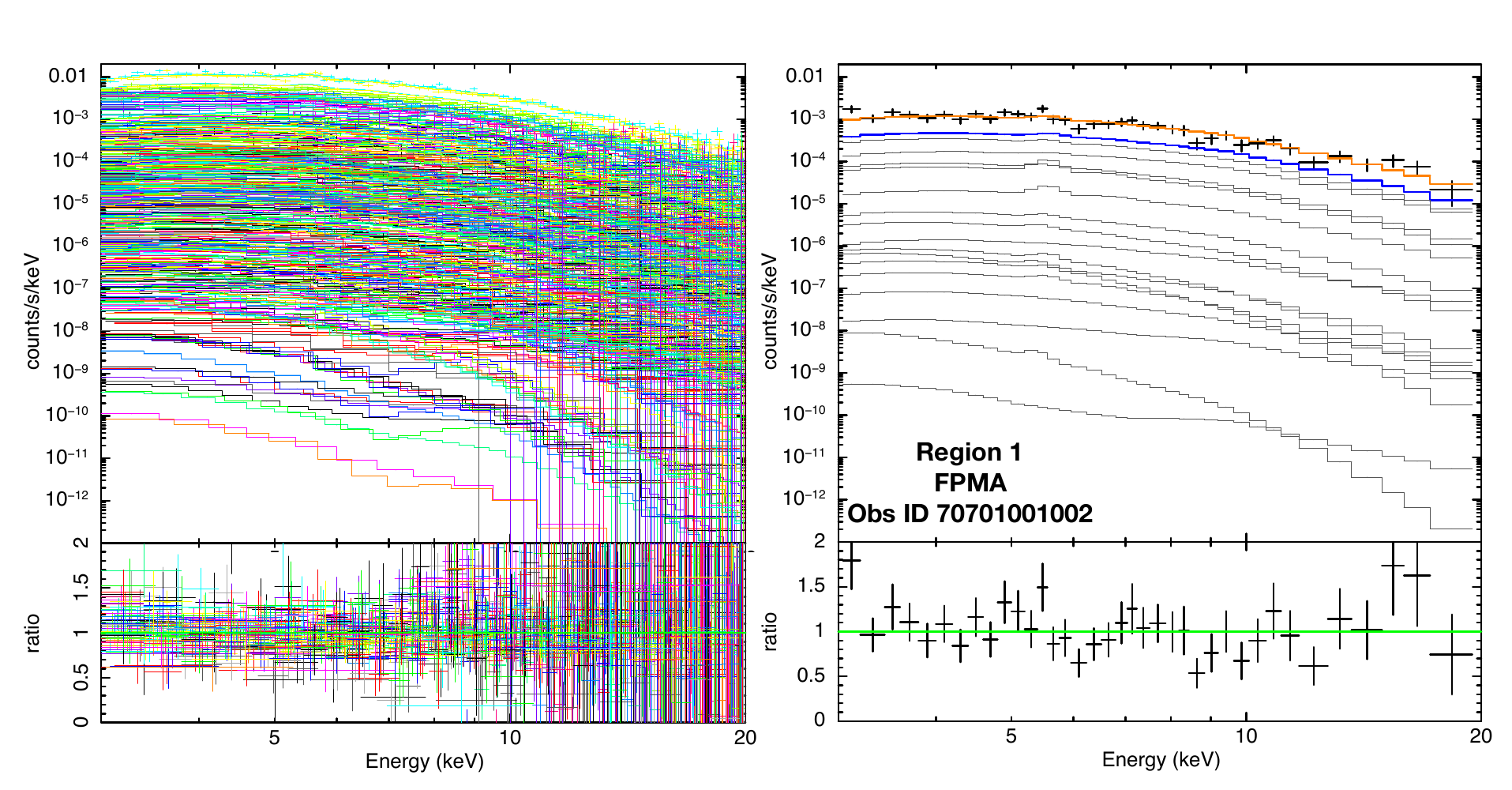}
\caption{Total cross-talk spectra for the AGN scenario final fit, (left panel) and the spectra for the AGN scenario, for Region 1, ObsID 70701001002 FPMA (right panel). In the right panel, black indicates the data, orange presents the total model, blue curve is the local contribution model from Region 1 and gray curves represent the scattering from other regions. \label{fig:crosstalkspectra}}
\end{figure*}

The final fitted spectra for the obscured AGN scenario is presented in the left panel of Fig.~\ref{fig:crosstalkspectra} as an example. This figure shows the spectra from each source, the total ARF curve and the individual contributions from all regions, for two observations and for both FPMA and FPMB: (1+1+17)~$\times$~2~$\times$~2 curves in total. In order to present a much less complicated version, we plotted the spectrum from Region 1, the total ARF curve and the individual contributions from all regions to Region 1 emission, for Obs ID 70701001002 and for FPMA as seen in the right panel of Fig.~\ref{fig:crosstalkspectra}

\begin{figure*}
\centering
\includegraphics[width=140mm]{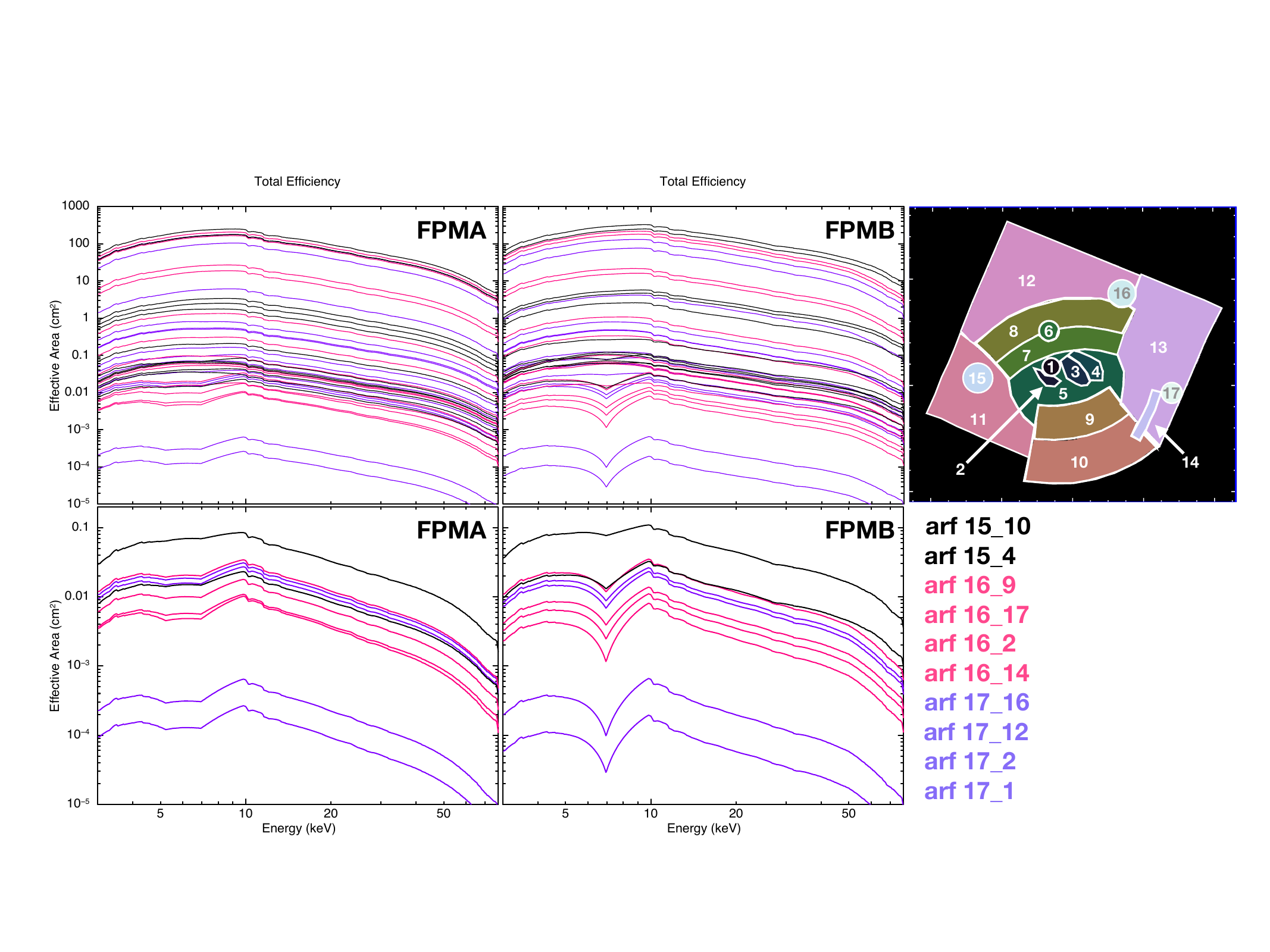}
\caption{All cross-ARFs for Region 15 (black), 16 (purple) and 17 (pink) from Obs ID 70701001002 (upper panel), and only the cross-ARFs of artifacts (lower panel), namely cross-ARFs: 15-10, 15-4, 16-2, 16-9, 16-14, 16-17, 17-1, 17-2, 17-12, and 17-16. \label{fig:arfdip}}
\end{figure*}

As seen in the {\tt nucrossarf} spectral fit of Region 1 spectra for ObsID 70701001002 FPMA, the shape of the weakest model seen in the right panel in Fig.~\ref{fig:crosstalkspectra} has an artifact- a dip within 5.0~-~10.0 keV. This model corresponds to the scattering from Region 17 to Region 1. We further investigated the spectra for similar artifacts and saw that point source crosstalk contributions to the regions that are far away from these sources show a similar behaviour. These curves are 15-10, 15-4, 16-2, 16-9, 16-14, 16-17, 17-1, 17-2, 17-12, 17-16, where the first numbers in the couples point to the origin of the scattered light, and the seconds correspond to where the scattered light reaches. These crossarfs with artifacts for Obs ID 70701001002 are plotted in the lower panel of Fig.~\ref{fig:arfdip}, where the upper panel shows the entirety of the ARFs. 

Since for the production of ARFs, the PSF images at various bands are used, these artifacts are bound to happen for faint point sources that are close to the edge of the FOV where the calibration for PSF is less than perfect. {\tt nucrossarf} samples the PSF images for extended sources therefore averaging out very small fluctuations. Therefore we do not see any similar artifacts from Region 14, Channel, which is also close to the edge of the FOV and has faint emission. At very low effective area, the level of the artifacts are comparable to data noise. We emphasize that, at this rate, ARF contributions where we see the dip artifact are effectively zero ($\sim$10$^{-9}$ counts/s) and the artifacts do not have an effect on any of our results.

\end{document}